\documentclass{aa}
\usepackage[varg]{txfonts}
\usepackage{url}
\usepackage{graphicx}
\usepackage{latexsym}
\usepackage{amsmath}
\usepackage{amssymb}
\usepackage{textcomp}
\usepackage{natbib}
\usepackage[labelfont=bf]{caption}
\usepackage{color}

\begin{document}

\title{Hubble imaging of V1331 Cyg: proper motion study of its circumstellar structures}

\author{A.~Choudhary\inst{\ref{inst1}} \and B.~Stecklum\inst{\ref{inst2}}
\and Hendrik Linz\inst{\ref{inst6}}}

\institute{
Institute of Mathematical Sciences, Chennai, India. \email{arpitac@imsc.res.in} \label{inst1}
\and
Th\"{u}ringer Landessternwarte Tautenburg, Sternwarte 5, D-07778, Tautenburg, Germany, \email{stecklum@tls-tautenburg.de}\label{inst2}
\and
Max Planck Institute for Astronomy, Heidelberg, Germany, \email{linz@mpia-hd.mpg.de}\label{inst6}
}

\abstract {}
{
The young star V1331 Cyg received previous attention because it is surrounded by an optical, arc-like reflection nebula. V1331 Cyg is commonly considered to be a candidate for an object that has undergone an FU-Ori (FUOR) outbreak in the past. This in turn could lead to a time-varying appearance of the dusty arcs that may be revealed by multi-epoch imaging. In particular, a radial colour analysis of the dust arcs can then be attempted to check whether the radial grain size distribution was modified by a previous FUOR wind.
}
{
Second-epoch imaging of V1331 Cyg was obtained by us in 2009 using the Hubble Space Telescope (HST). By comparing this to archival HST data from 2000, we studied the time evolution of the circumstellar nebulae. After a point spread function subtraction using model point spread functions, we used customised routines to perform a proper motion analysis. The nebula expansion was derived by deconvolving and
correlating the two-epoch radial brightness profiles. Additional data from other facilities -- TLS, UKIDSS, SPITZER, and HERSCHEL -- were also incorporated to improve our understanding of the star in terms of environment, viewing angle, bipolar outflow length, and the FUOR phenomenon.
}
{
The outer dust arc is found to be expanding at $\approx$14.8$\pm$3.6\,km\,s$^{-1}$ on average. The expansion velocity for the inner ring is less consistent, between 0.8\,km\,s$^{-1}$ and 3.0\,km\,s$^{-1}$. The derived radial colour profiles do not indicate a spatial separation of the dust grain sizes. The {\sl HERSCHEL} 160\,$\mu$m images show for the first time thermal emission from dust probably residing in the outer arc. By viewing V1331 Cyg almost pole-on, the length of the bipolar outflow exceeds previous estimates by far.
}
{
The outer arc expansion timescale is consistent with the implantation time of the CO torus, which supports the hypothesis of an outburst that occurred a few thousand years ago. The azimuthal colour variation of the outer arc is probably due to changes of the scattering angle, imposed by a tilt or helical geometry of the dust configuration.
}

\maketitle

\section{Introduction}
Massive molecular cloud cores are the birth place of stars. A gravitationally unstable cloud core breaks down into fragments, giving birth to several stellar systems. Each fragment, 
eventually harbouring a protostar,
develops an accretion disk and an envelope around it. Matter falls onto the disk from the envelope. The disk transports the matter farther inwards, finally accreting it onto the protostar. During this phase, the disk may develop thermal and gravitational instabilities, leading to occasional enhanced accretion for a short duration. The FU Orionis phenomenon, which is observed among a few young stars, might be related to this phase. We refer to the review by \citet{1996ARA&A..34..207H}. FU Orionis-type objects (herafter FUORs) are named after FU Orionis, a star in Orion that rose in optical brightness by a factor of 6 magnitudes in a period of six months in 1936 \citep{1966VA......8..109H, 1977ApJ...217..693H}. Some other sources have been identified since then whose proporties are similar to those of FUORs (IR excess from circumstellar dust, P~Cygni profiles in Balmer lines, near-IR CO overtone absorption).

Most notable among the FUORs that have been detected since 1936 is V1057 Cygni. The pre-outburst spectrogram of V1057 Cygni resembles that of a T Tauri star \citep{1958ApJ...128..259H}. \citet{1985ApJ...299..462H,1987ApJ...312..243H} have explained FUOR properties in relation to accretion through circumstellar disk onto a T Tauri star. Outbursts occur when the mass-accretion rate from the disk onto the star increases by orders of magnitudes. In addition, optically bright T Tauri stars are the most evolved 
young stellar objects (YSOs)
that are associated with Herbig Haro objects (HHOs) or molecular outflows \citep{1998AJ....116..860M}.

\object{V1331 Cyg} \mbox{(RA=21:01:09.210, Dec=+50:21:44.77(J2000))} is a young star whose visual extinction is A$_{\rm v}\sim2.4$ \citep{1993AJ....106.2477M}. It is located at the border of the dark cloud LDN~981. \citet{1964ApJ...140.1409K} found an arc-like reflection nebula surrounding V1331 Cyg.  The reflection nebula around V1331 Cyg is similar to those found around FUORs  \citep{1987PASP...99..116G},  which supported the assumption that the star is
a FUOR. The distance to V1331 Cyg was estimated to be in the range of 694\,pc \citep{1981A&A...101..105C} to 550\,pc \citep{1991SvA....35..135S}. We use 550\,pc as the distance for our calculations here. 

The strong P~Cygni absorption lines emphasise on the presence of circumstellar matter surrounding the star \citep{1981A&A...101..105C,1984ApJ...280..749M}. A molecular outflow with a mass-loss rate of $1.0\times10^{-7}\,M_{\odot}$ per year was found by \citet{1988ApJ...330..897L}. On the basis of submillimetre continuum observations, \citet{1991ApJ...382..270W} suggested that there might be a compact circumstellar disk. CO synthesis maps by \citet{1993AJ....106.2477M} provided additional evidence that a massive disk of $0.5\,M_{\odot}$ exists  around
V1331 Cyg that
is surrounded by a flattened gaseous envelope. \citet{1998AJ....116..860M} found the object to be associated with HHOs and suggested there
might be a 
wiggled
jet. The total spatial extent of the bipolar outflow is expected
to be about 1 pc, as suggested by them. 
Based on archival HST WFPC2 imaging,
\citet{2007ApJ...656..287Q} showed that the star has two nested rings of $\sim$9000 and $\sim$3300 au radii, respectively. The
authors also claimed that the incomplete outer dust ring section to the NW is caused by the dark cloud LDN 981 that is located in the foreground. An additional dust arc can also be seen in the SW direction. 

\citet{2011ApJ...738..112D} suggested based on high-resolution IR spectroscopy of OH and $\mathrm{H}_2\mathrm{O}$ that the source
is seen from an almost pole-on viewing angle. 
Based on the detection of
high-velocity absorption lines of metals (Fe II 5018, Mg I 5183, and K I 7699), which form in post-shock gas in the jet, \citet{2014MNRAS.442.3643P} concluded that V1331 Cyg is seen through its jet. 
Since atomic or ionic jets have a small opening angle of only
a few degrees, see \citet{2004A&A...415..189M}, this implies an almost pole-on view.

The dust arcs seen today could be a result of ejections in a past FUOR outburst, or they might be due to a pole-on view, therefore
we might be seeing scattered light from the remnants of the outflow cavity.
What we observe is the projection of a three-dimensional structure onto the plane of the sky that appears as an arc. In either case, the dust arcs are expected to be expanding as they move away from the source. With our second-epoch observational data for V1331 Cyg from HST, the prime aim is to compare the two epochs of data to measure the inner and outer dust arc radial expansion. 

During the outburst, the short-lived strong wind enforces a ram-pressure on ambient dust grains. 
Since the ratio between cross-section and grain mass is reciprocal to the grain size, smaller grains experience higher acceleration than larger ones \citep{1977ApJ...217..693H,2016arXiv160302899H}. 
This leads to a spatial separation of dust particles according to size if the wind subsides quickly enough.
Since the dust scattering
depends on grain size, a radial gradient in particle size would imprint itself as a colour gradient. 
We analysed the outer dust ring colour to check whether the particle
separation hypothesis is plausible. 

We use additional data from UKIDSS, SPITZER, TLS and {\sl HERSCHEL}/PACS in this paper to investigate the circumstellar environment of the source. The bipolar outflow length and inclination angle are also studied and the results are updated. We use the terms dust ring and dust arc interchangeably throughout.

\section{Observations}
\subsection{Hubble data}
V1331 Cyg was observed twice with the WFPC2 camera of the HST, with a time gap of almost ten years. 
The WFPC2 handbook by \cite{2004WFPC2} describes the HST WFPC2 specifications, and Table~\ref{v1331} gives details of the observations. 

\begin{table*}[ht]
\caption{V1331 Cyg observations with HST-WFPC2}
\centering
\begin{tabular}{c c c c}
\hline \hline
\noalign{\vskip 1mm}  
Data set & Filter & Exposures & Instrument\\ [0.5ex]
\hline \noalign{\vskip 1mm} 
& F606W & 3 exposures for 10, 230, 230 s each & PC,WF3 \\[-1ex]
\raisebox{1.5ex}{First epoch (2000)} & F814W & 2 exposures for 10, 100 s each & PC \\[1ex]

& F450W & 4 exposures for 7, 7, 2300, 2300 s each & WF3 \\[-1ex]
\raisebox{1ex}{Second epoch (2009)} & F606W & 3 exposures for 3, 1100, 1100 s each & WF3 \\
& F814W & 3 exposures for 3, 1100, 1100 s each & WF3 \\
\hline
\end{tabular}
\label{v1331}
\end{table*}

We downloaded the HST pipeline products from the MAST archive. The data were bias and drizzle corrected. Using the image reduction and analysis facility (IRAF), the WFPC2 images were cleaned from hot pixels and cosmic ray defects. Two images from the same filter with the same exposure time were combined using the IRAF imcombine task. The process was repeated separately for images from both epochs in all filters. The {\sc Crreject} feature was used to clean the data from cosmic defects in both epochs.

\subsection{UKIDSS survey}
We retrieved UKIDSS (UKIRT Infrared Deep Sky Survey) 
infrared, JHK-band images of V1331 Cyg from the WFCAM (Wide Field Camera) science archive. The pipeline processing and science archive are described in \citet{2008MNRAS.384..637H}. The survey instrument WFCAM has four $2048 \times 2048$ Rockwell Hawaii-II PACE arrays at $94\%$ spacing.  The pixel scale of $0\farcs4$ gives an exposed solid angle of 0.21 sq.degrees. 
As a result of   microstepping, the image sampling amounts to $0\farcs2$.
Observation details are given in Table~\ref{ukirt}.

\begin{table*}[ht]
\caption{V1331 Cyg observations with UKIDSS in 2010}
\centering
\begin{tabular}{c c c c c}
\hline \hline \noalign{\vskip 1mm} 
Broadband filter & Mean $\lambda\,[\mu \rm{m}]$ & Exposure time [s] & FOV & Instrument \\[0.5ex]
\hline \noalign{\vskip 1mm} 
J  & 1.25 & 10 & $\approx$27$'\times 27'$ & WFCAM \\
H & 1.63 & 10 & $\approx$27$'\times 27'$ & WFCAM \\
K & 2.2   &   5 & $\approx$27$'\times 27'$ & WFCAM \\ 
\hline
\end{tabular}
\label{ukirt}
\end{table*}

\subsection{TLS observations and SPITZER imaging}
For the bipolar outflow structure analysis, optical imaging of V1331 Cyg was performed using the 2m Tautenburg telescope in its Schmidt configuration, offering a free aperture of 1.34\,m, in April 2003 and November 2011. On both occasions, I, H$\alpha$, and [SII] filters were used. The images were processed to account for flatfield variations and bad pixels using our custom IDL routines, and they were astrometrically calibrated by means of the astrometry.net code \citep{2010AJ....139.1782L}. For each filter, the frames were registered and stacked, yielding total images with integration times of 12, 80, and 100 minutes. 

HHO spectroscopy was done using Nasmyth Spectrograph at the 2m Tautenburg telescope. The final spectra for the southern and northern HHOs were obtained with a total integration time of 9600\,s and 6000\,s, respectively. Data from two nights in July 2013 were added, accounting for the change of the projected Earth velocity towards the target. Spectral lines originating from airglow in the upper atmosphere served for wavelength calibration. To convert the pixel coordinates into wavelength, a dispersion relation was established by fitting the actual sky lines using a least-squares minimisation. We applied the  optimal extraction technique to obtain the spectrum.

The SPITZER IRAC images were downloaded from IRSA (Infrared Science Archive). The instrument description is detailed in \citet{2004ApJS..154...10F}. Observation with AORKEY 27067904 
(principal investigator Tim Spuck) 
was performed on July 18, 2008. 
The individual frames have an exposure time of 10.4\,s each, and a total time of about 200\,s was spent per filter to create the mosaic image.

\subsection{HERSCHEL observations}\label{Subsec:HERSCHEL-obs}
V1331 Cyg was observed with the {\sl {\sl HERSCHEL}} spacecraft \citep{2010A&A...518L...1P} within the Open-Time Project OT1\_jgreen02\_2 \citep[cf.][]{2013ApJ...772..117G}. The related observations with the bolometer cameras of {\sc PACS} \citep{2010A&A...518L...2P} took place on July 22, 2011 (operational day 800) and were contained in the observational IDs 1342225252--55. The PACS prime mode was employed with a nominal scanning velocity of 20$''$/second. Data were taken in all three PACS filters (70, 100, and 160~$\mu$m) in the mini-map mode and resulted in maps with a useful field
of view of roughly 3.5 arcmin. 
For the 160~$\mu$m data, roughly 160~s of dwell time are accumulated per map pixel in the middle
of the combined map. For the two other filters, it is half of that.
We re-reduced the existing {\sl {\sl HERSCHEL}} archive (HSA) data, starting from the Level1 data. These we retrieved from the currently running bulk processing with the Scientific Product Generation (SPG) 13.0 software version. The pointing products of this new edition of the data include the newly developed data reduction step {\sc calcAttitude}: a correction of the frame pointing product based on the {\sl {\sl HERSCHEL}} gyroscope house-keeping \citep[see e.g.][]{2014ExA....37..453S}. This often improves the absolute pointing accuracy, but more importantly, it also mitigates the pointing jitter effect on individual frames. We used these gyro--corrected frames and performed the de-striping and removal of 1/f noise using {\sc JScanam} within the {\sl {\sl HERSCHEL}} Interactive Processing Environment ({\sc HIPE}), version 13.0 b5130. {\sc JScanam} is the {\sc HIPE} implementation of the popular data reduction scheme for {\sl {\sl HERSCHEL}} data named {\sc Scanamorphos} \citep{2013PASP..125.1126R}. For the individual detector pixel distortion correction and the final mapping, the implementation of the drizzle algorithm \citep{2002PASP..114..144F} within {\sc HIPE,} called {\sc photProject,} was used, which allows experimenting with different pixel fractions.

Reference point spread functions (PSFs) were taken from the PSF tarball provided by the {\sc PACS} instrument team\footnote{\url{ftp://ftp.sciops.esa.int/pub/hsc-calibration/PACS/PSF/PACSPSF_PICC-ME-TN-033_v2.0.tar.gz}}. For further analysis, these were rotated by 13.8\degr{}  counterclockwise to align them with the orientation of the {\sc PACS} PSF during the V1331 Cyg observations.

\section{Results}

\begin{figure}[ht]
\begin{center}
\includegraphics[width=9cm, height=9cm]{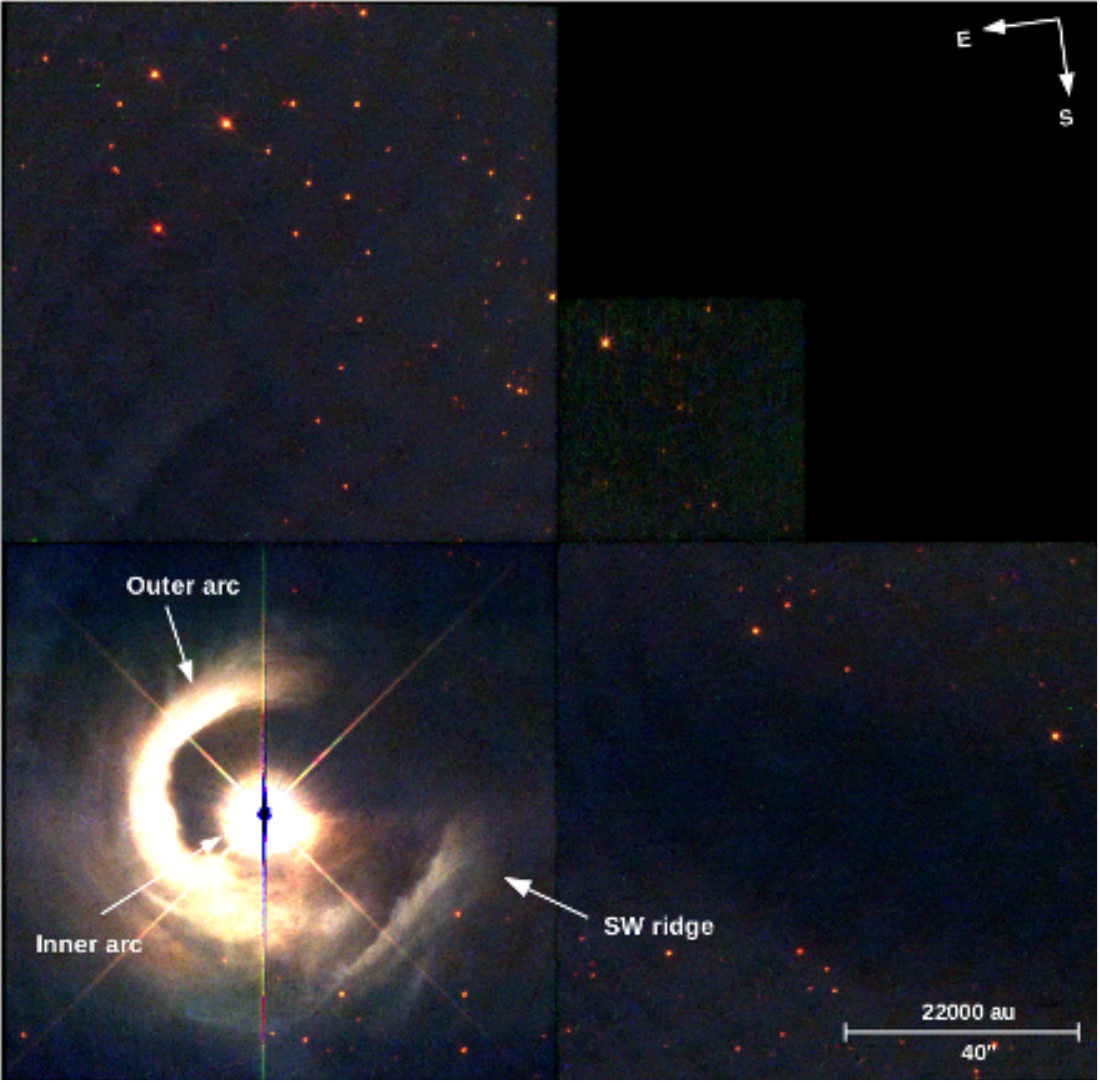} 
\end{center}
\caption{Second-epoch full-field colour  WFPC2 image of V1331 Cyg, showing the arc structure. 
Details of the scattering nebula and the adjacent dark cloud are obvious. We note the straight division between the lit (south) and dark (north) areas west of the star.
}
\label{color}
\end{figure}

To understand the origin and nature of the dust arcs, that is, to check the validity of a possible arc expansion, we performed a proper motion analysis on the outer and inner dust arcs. The most simple case of sole radial expansion in the sky plane is assumed here. 
While this approach represents a rough approximation of the actual proper motions,
it provides the advantage of focusing all the image information on one parameter. For this purpose,
radial profiles from both epochs were calculated and compared. The arc structure of V1331 Cygni is shown in a colour composite in Fig.~\ref{color}. 
The missing section to the north-west, which would  complete the arc to a ring, can be easily identified. Its southern boundary runs straight almost east-west and is reminiscent of a shadow. This feature is  important for the discussion of whether the missing section is caused by LDN\,981 (cf. Sect. \ref{ldn}).

\subsection{Radial dust arc profiles: Proper motion analysis}  \label{radial_analysis}

The method used to obtain radial profiles of the outer dust arc is as follows.

We used F606W-filter images from the two epochs, cleaned from bad pixels and cosmics. The two data sets were processed with
custom 
IDL\footnote{IDL is a trademark of Harris Corp.}
routines. 
Background light was differently scattered on wide-field camera detectors 3 and 4. To correct for this effect, the affected third-quadrant pixels of the image frame were replaced by the
azimuthal median of pixels in the fourth quadrant. Model PSFs generated with the Tiny Tim standalone application \citep{2011SPIE.8127E..16K} were then subtracted from the images. 
Tiny Tim required an input for the
spectrum of the star, which was obtained from the
Vizier photometry viewer. Images were cropped to form square arrays. The images from first and second epoch had different roll angles, which led to mutual field rotation on the
CCD detector. The transformation matrix between the pixel coordinates of the two epochs was derived from field stars detected in the full WFPC2 frames.
This was done iteratively to avoid the influence of stars with substantial proper motion. 
In the second step,
10 out of the initial 11 stars were eventually used to calculate the transformation parameters, which are listed in Table~\ref{trans}.

\begin{table}[h]
\caption{List of parameter values for the outer dust arc to map second-epoch images with respect to first-epoch images.
}
\centering
\begin{tabular}{l c}
\hline \hline \noalign{\vskip 1mm}
Parameter & Transformation value \\
\hline \noalign{\vskip 1mm}
Xshift $\pm$ error & $-135.5907 \pm 0.1431$ [pixel]\\
Yshift $\pm$ error & $235.4284 \pm 0.1431$  [pixel]\\
Rotation $\pm$ error & $0.51745 \pm 0.0031$ [radian]\\
Stretch $\pm$ error & $1.00006 \pm 0.0045$ \\
Reduced $\chi^2$ & $0.055171$ \\
\hline
\end{tabular}
\label{trans}
\end{table}

The second-epoch image was remapped using values from Table~\ref{trans} to match  the first-epoch image.

\begin{figure*}[ht]
\centering
\includegraphics[clip,width=9cm]{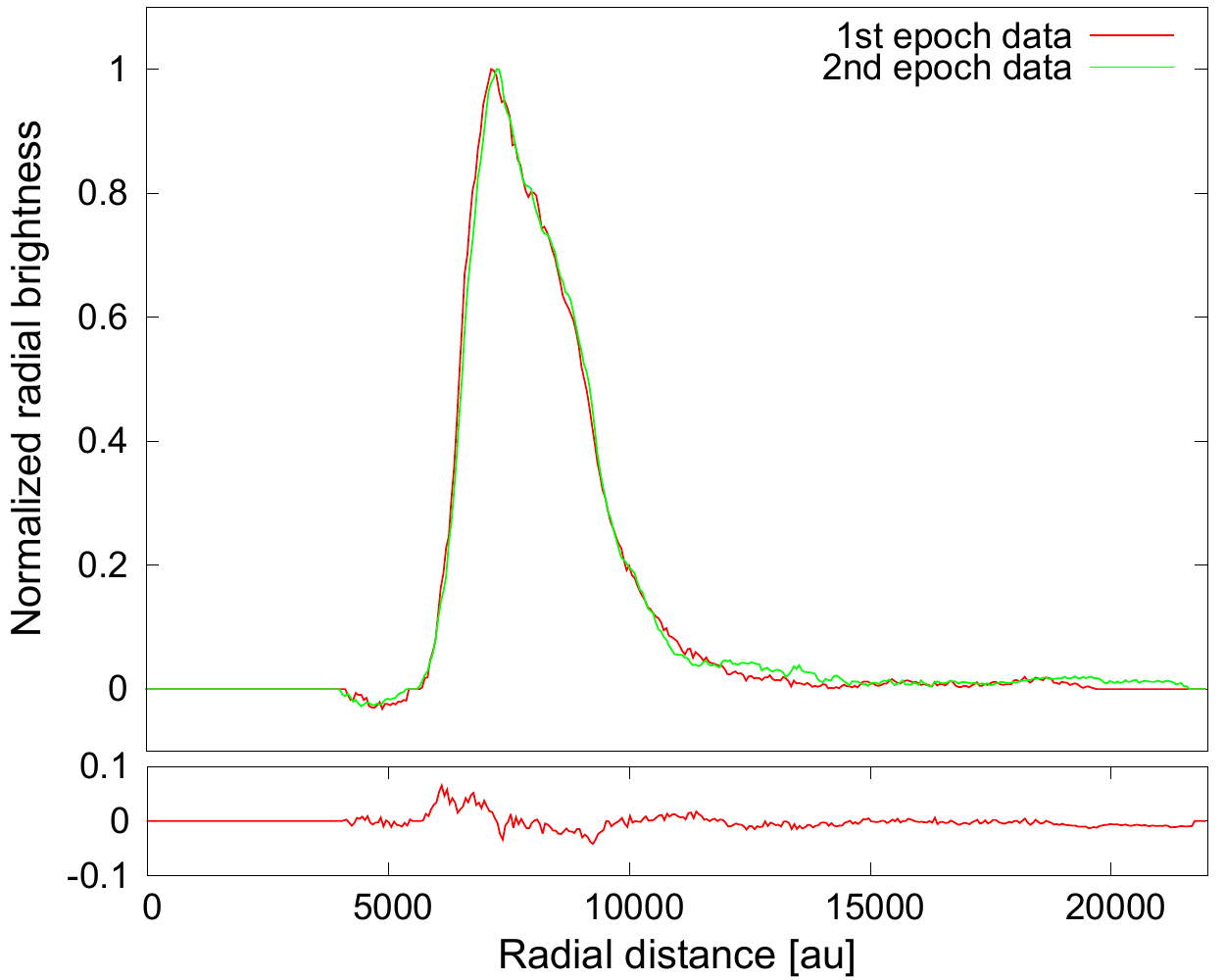}
\includegraphics[clip,width=9cm]{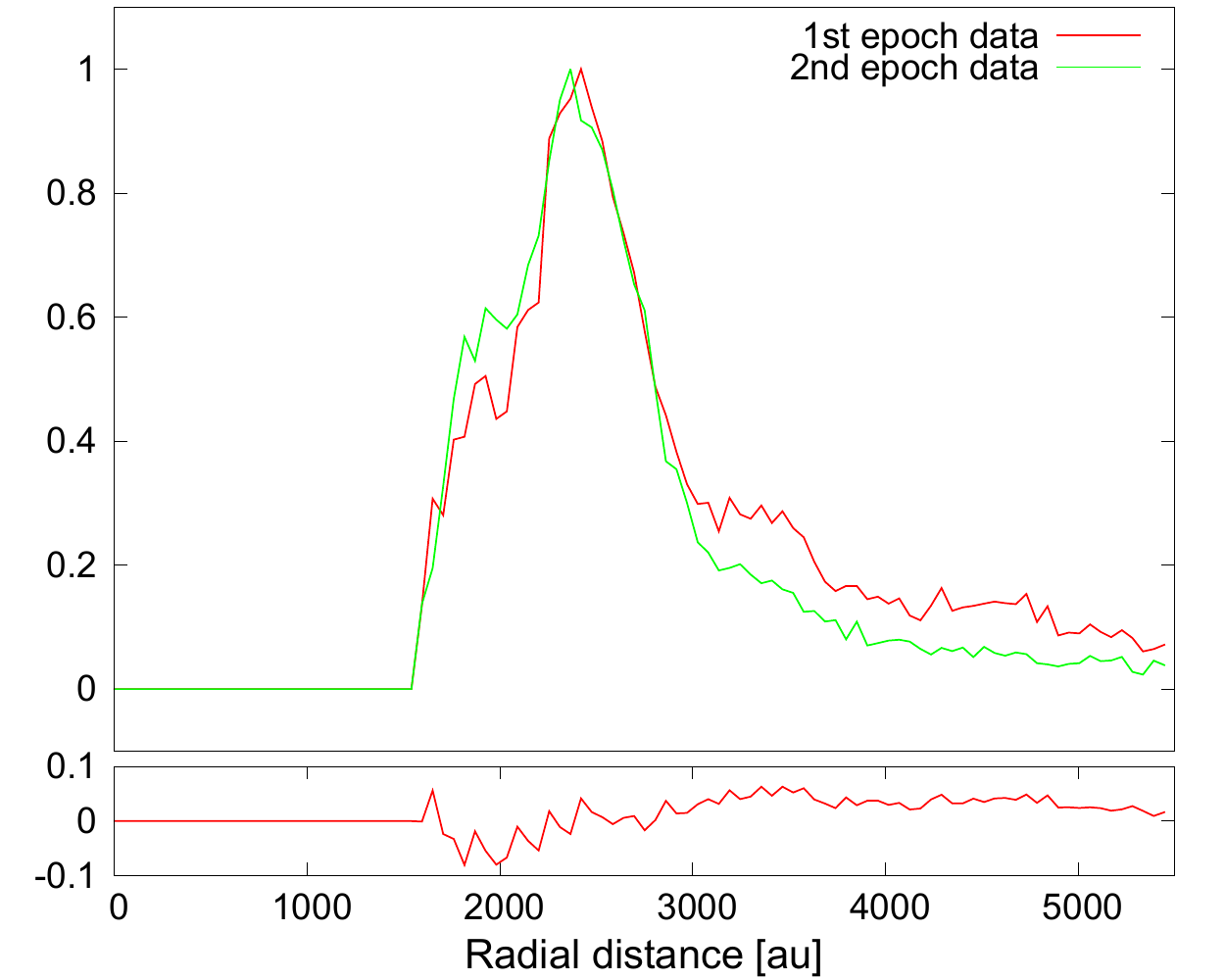}
\caption{Radial outer (left) and inner (right) dust arc profiles of V1331 Cyg for the F606W filter image from both epochs are over-plotted for comparison. The lower panels show the difference in profiles.
}
\label{radial}
\end{figure*}

We then 
derived the radial-brightness profiles for the registered frames of the two epochs and derived their mutual shift. 
Scattered stellar light was removed from both images in the same fashion before obtaining the radial profiles. Since the arc centre does not coincide with the stellar position, it was determined as outlined in the following.
Using an initial guess for centre of the outer arc 
as pixel coordinate origin,
images were converted from Cartesian into polar coordinates. The azimuthal median for each radial value was obtained from polar profiles for a range of 5\degr$\dots$220\degr{} in position angle (PA), which gave us radial profiles of the dust arc.
The final outer dust arc centre was obtained by using Powell minimisation \citep{1964CompJ} to determine the centre shift that yields the most narrow profile. The median radial dust arc profiles are shown in Fig.~\ref{radial} (left).


The next task was to determine whether there is any measurable shift between profiles from the two epochs. To this aim, two methods were applied.

The first method is 
based on the cross-correlation lag, which indicates that the second-epoch profile is 0.64 pixel farther away from the star. This result is very robust against various levels of oversampling (ranging from 20 to 100) and requires no zeroing of the outer profile regions. 
It corresponds to an average expansion velocity of 16.6\,km\,s$^{-1}$.
The second method to retrieve the mutual shift employs the deconvolution of the radial profiles. For this purpose, the profiles were oversampled by a factor of 100, and the first-epoch profile served as deconvolution kernel. 
The initial application to the full profiles yielded a large shift that should have been obvious, but this was not the case.
The deconvolution is apparently sensitive against differences in the outer plateau of the profiles. For this reason, only their cores were kept while the plateau range (outside 12000\,au, cf. Fig.\,\ref{radial} left) was discarded.  
Thereby, a shift of 0.50\,pixel was found, corresponding to an
expansion velocity of 13.0\,km\,s$^{-1}$. An attempt was made to derive a formal error using Monte Carlo realisations of the deconvolution on a smoothed second-epoch profile with white noise added. However, given the systematics induced by the choice of the outer profile cut-off, we refrained from considering this noise estimate as representative. Thus, in the absence of a rigorous error treatment, we adopted the difference between the two results as a measure of the uncertainty, that is, 0.14 pixel or 3.6\,km\,s$^{-1}$, and used their average of 0.57 pixels or 14.8\,km\,s$^{-1}$ as estimate for the overall tangential expansion velocity for the outer arc.



The inner arc radial profiles were obtained using only individual WF3 frames instead of the full WFPC2 frames. This was done because
the 
inner arc is much closer to the central source and needed a better PSF subtraction. 
For this purpose, the HST archive was searched for F606W images of stars that match V1331 Cygni in both brightness and colour
(data kindly provided by K.~Stapelfeldt).
The transformation parameters calculated for the individual WF3 frames are given in Table~\ref{in_trans}. The 
centre of the
outer arc is slightly offset from the centre 
of the inner arc. The inner dust arc radial profiles were obtained using the same method as for 
the
outer dust arcs. Figure~\ref{radial} (right) shows the resulting inner arc radial profiles.

\begin{table}[ht]
\caption{List of parameter values for the inner dust arc to map the 
second-epoch WF frame with respect to the 
first-epoch WF frame.}
\centering
\begin{tabular}{l c}
\hline \hline \noalign{\vskip 1mm}
Parameter & Transformation value \\[0.5ex]
\hline \noalign{\vskip 1mm}
Xshift  $\pm$ error  & $-81.4429 \pm 0.7883$ [pixel]\\
Yshift $\pm$ error & $215.1916 \pm 0.7891$ [pixel]\\
Rotation $\pm$ error & $0.51748 \pm 0.00469$ [radian]\\
Stretch $\pm$ error & $0.99777 \pm 0.03511$ \\
Reduced $\chi^2$ & $0.067695$ \\ 
\hline
\end{tabular}
\label{in_trans}
\end{table}

The inner arc profiles are not very well constrained because of their proximity to central source and residuals after PSF subtraction. Using the
deconvolution method, we calculated a shift of 0.03\,pixel. By again running Monte Carlo simulations, the shift obtained was about 0.13\,pixel. The two pixel shift values correspond to expansion velocities of about 0.8\,km\,s$^{-1}$ and 3.4\,km\,s$^{-1}$ .

\subsection{HST F606 image subtraction}
\begin{figure}[h]
\begin{center}
\includegraphics[trim={7cm 2.5cm 7cm 5cm},clip,width=8cm]{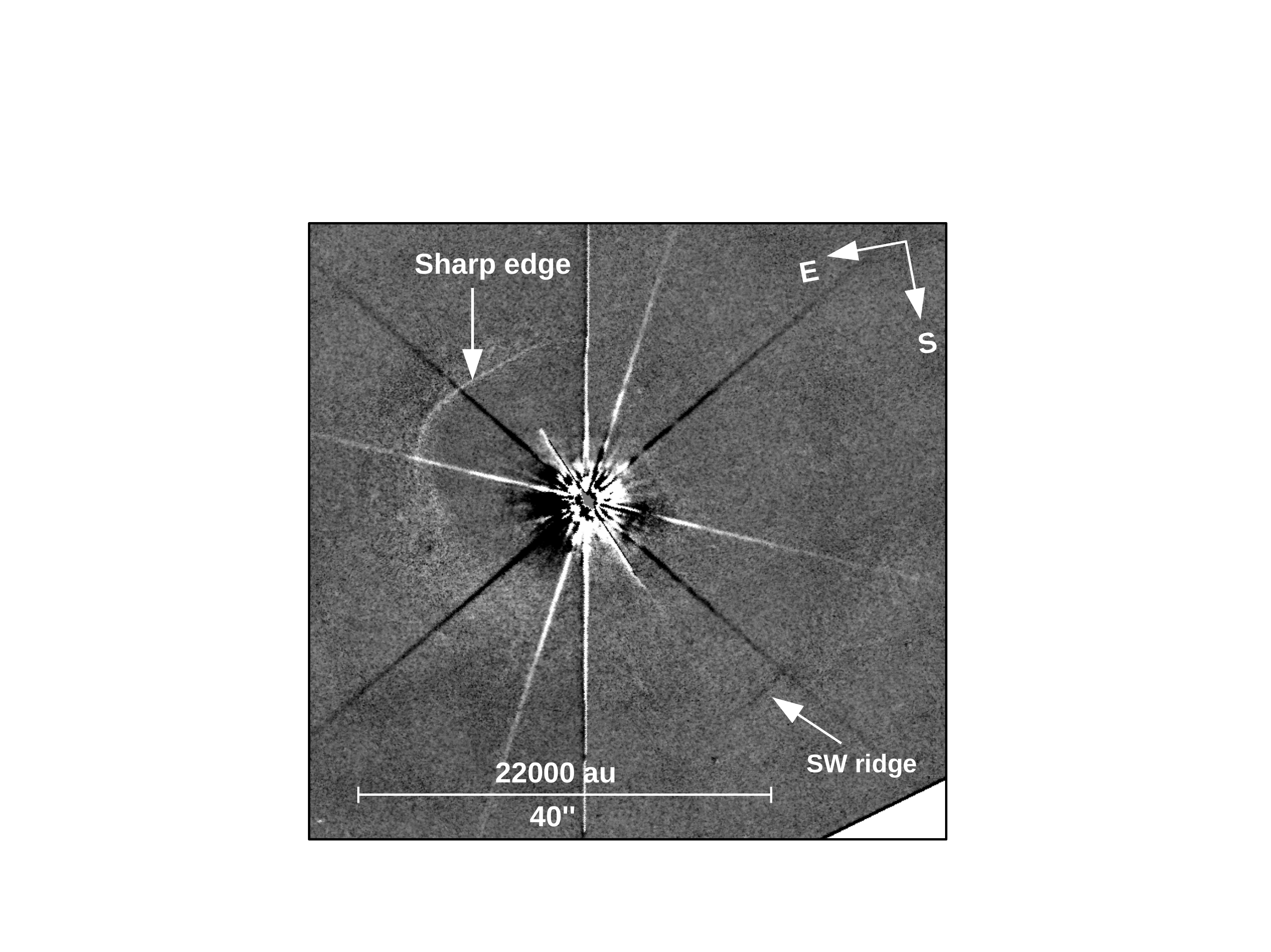}
\caption{Difference image 
(inverse grey scale)
of the F606W first- and second-epoch frames, showing signs of expansion in the NE part of the outer ring.}
\label{diff_im}
\end{center}
\end{figure}

We subtracted the first-epoch F606W image straight away from the second-epoch image
to identify differences of nebulosity and traces of radial expansion in the outer arc. Although we found no brightness variations as such, we found indications
for outward motion of the outer arc inner boundary in the north-east (shown in Fig.~\ref{diff_im}). There were also residuals from the subtraction at the location of the SW ridge, but with the opposite sign. Both features might form if the coordinate transformation were affected by star(s) with substantial proper motion that
would lead to a small shift, but this has already been taken care of. This finding
indicates radial expansion between the two epochs and thus supports our radial profiles analysis. The inward motion of the SW ridge is probably part of a more complex kinematics. 

\subsection{Radial dust arc profiles: Particle separation hypothesis}  \label{variable}

\begin{figure}[h]
\includegraphics[clip,width=9cm]{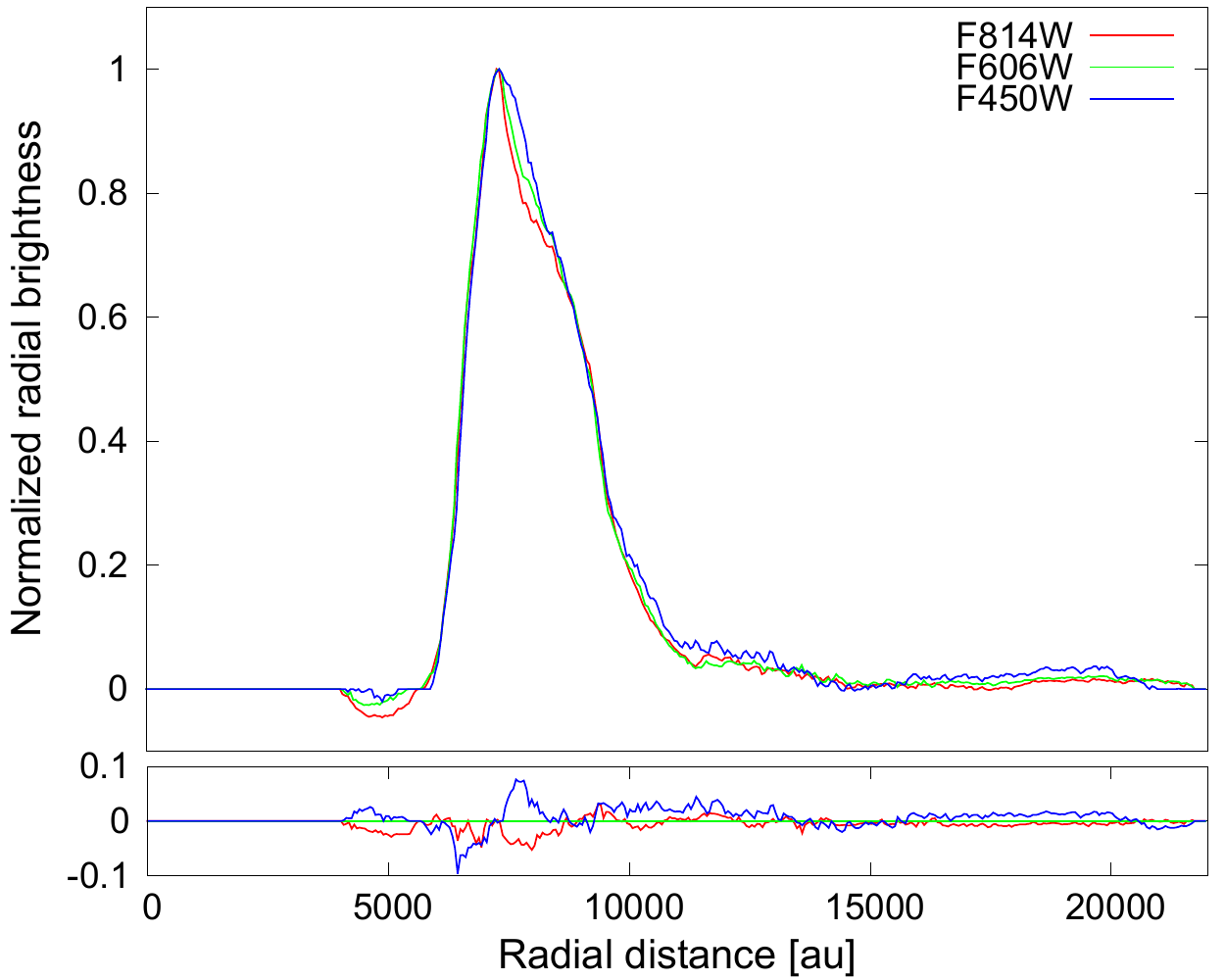}
\caption{Second-epoch normalised outer dust arc radial profiles obtained in all three filters. The lower panel shows the comparison of profiles with respect to  the
F606W filter. }
\label{particle_sep}
\end{figure}

To test the particle separation hypothesis, that is, to determine
whether the dust grain size distribution changes across the outer arc as a result of the grain-size-dependent acceleration by a possible short-lived FUOR wind,
we obtained and analysed the outer arc radial profiles from second-epoch images for all three filters (F450W, F606W, and F814W). The same procedure as detailed in Sect.~\ref{radial_analysis} was used. The resulting plot of the normalised profiles is shown in the
upper panel of Fig.~\ref{particle_sep}.

Any major profile variations are difficult to discern from Fig.~\ref{particle_sep}, upper panel, so the difference plots are shown in Fig.~\ref{particle_sep}, lower panel, for a better comparison.

\begin{figure}[ht]
\centering
\includegraphics[trim={4cm 0.4cm 4cm 1cm},clip,width=9cm]{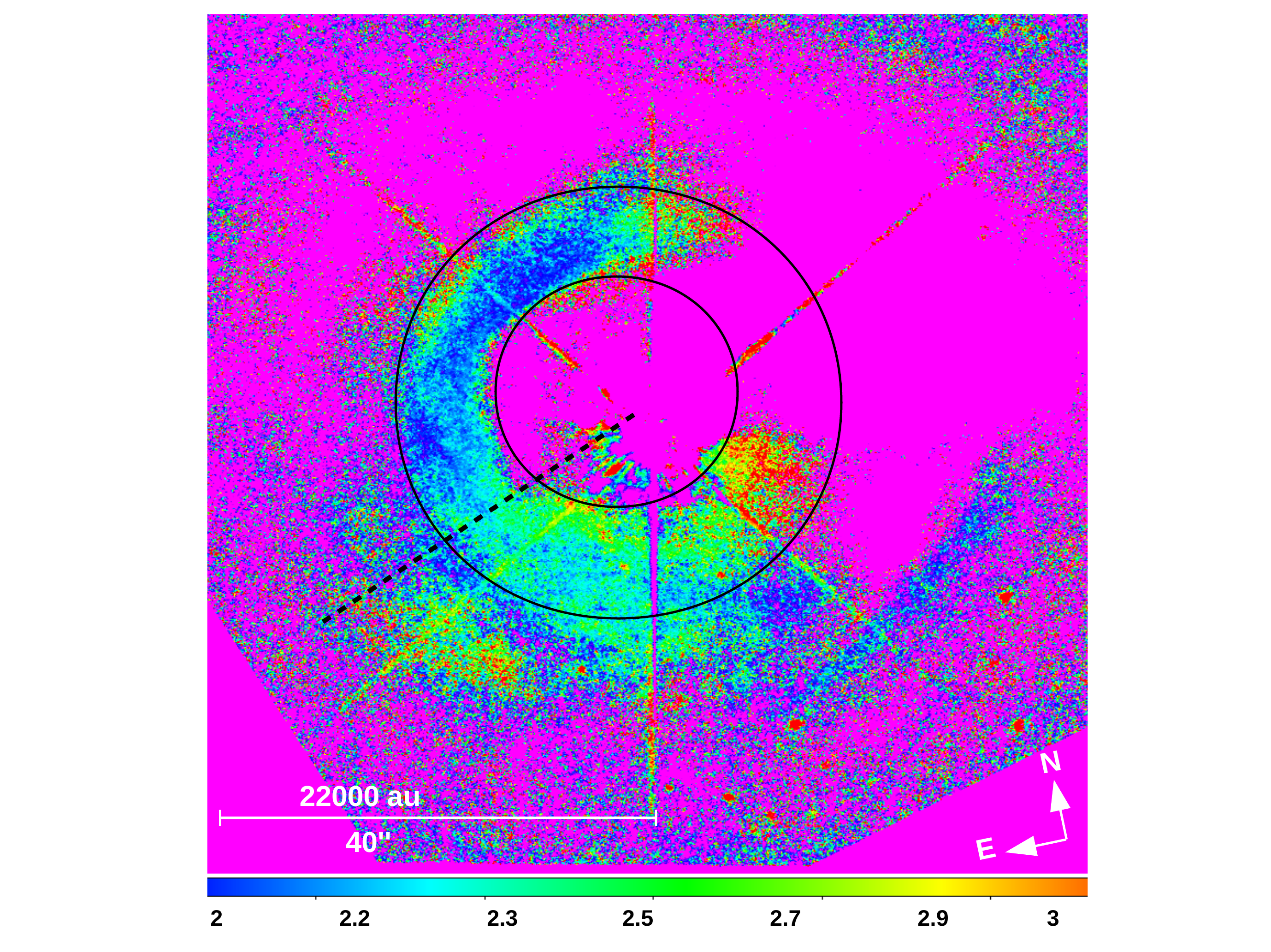}
\caption{The [F450W-F814W] colour index image from the
2nd epoch frames. 
The lower bar indicates the colour index range [mag],
and the dashed line denotes partition between two parts of the outer arc. 
The azimuthal colour profile as well as the brightness-colour relation are based on the annular region within the two circles.
}
\label{f4_8}
\end{figure}

The profiles match well in most parts
except for the radius values between $\sim$7000-9000\,au. This region covers exactly the central part of the dust arcs. This ignores any local and random variations in the
dust arcs along the
azimuthal range. It is hence difficult to judge whether the slight variation seen in the
normalised arcs is  dominated by the properties of local features situated at particular locations in the arc, or whether it is an equally distributed effect in dependence of the radial distance to the centre.
To investigate that, we calculated the [F450W-F814W] image shown in Fig.~\ref{f4_8}. The
azimuthal range from north to east appears bluer
than the remaining dust arc. The colour gradually gets more red as we move towards the south-west diffraction spike, starting from the
east.

For what concerns the outer range of the profile (distances $\gtrsim$15000\,au), the normalised brightness is slightly bigger the shorter the wavelength. This could be an indication of an enhanced fraction of smaller grains compared to the dust population of the core profile, possibly due to mixing with ISM dust. 

\begin{figure*}[ht]
\includegraphics[clip,width=9.1cm]{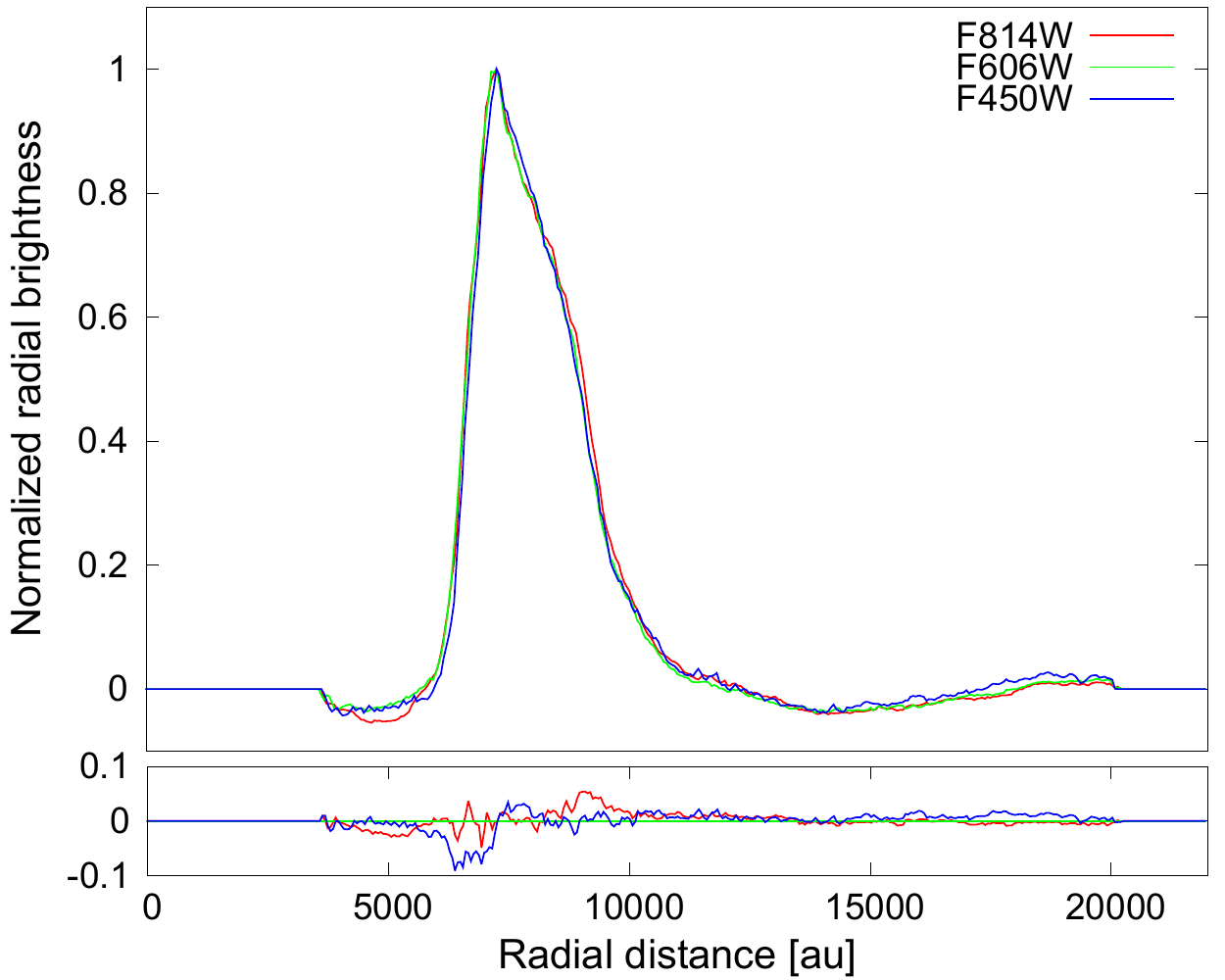}
\includegraphics[width=9.1cm]{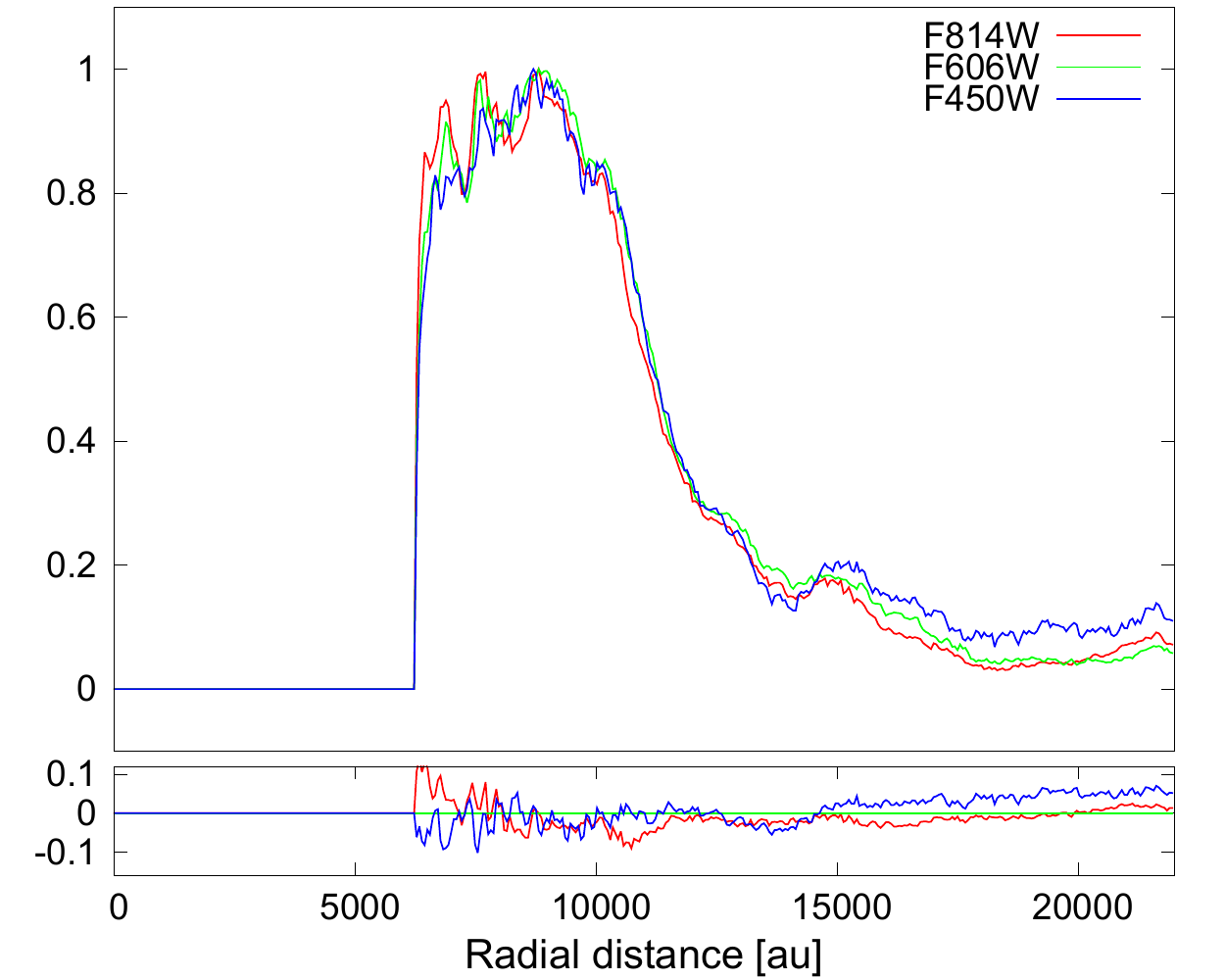}
\caption{Radial profiles of the outer dust arc for the azimuthal ranges extending from north to the east (left), and from east to the SW spike (right). The lower panels show the difference in profiles.}
\label{color_sep}
\end{figure*}

\subsection{Photometry of V1331 Cyg}
V1331 Cyg is saturated in long-exposure HST-WFPC2 images.
The photometry of the central source was derived from
the unsaturated images, which have short exposure times of a few seconds. Table~\ref{unsaturated} summarises the results.

\begin{table*}[!t]
\caption{V1331 Cyg photometry}
\centering
\begin{tabular}{c|c c c|c c c}
\hline \hline
& \multicolumn{3}{c}{Brightness [mag]} & \multicolumn{3}{c}{Colour index [mag]} \\[0.5ex]
\hline
& F450W & F606W & F814W & F450W-F606W & F450W-F814W & F606W-F814W \\[1ex]
\cline{2-7}
1st epoch & - & $11.98 \pm 0.003$ & $10.93 \pm 0.01$ & - & - & $ 1.053 \pm 0.01$ \\[1ex]
\hline
2nd epoch & $12.83 \pm 0.008$ & $11.77 \pm 0.005$ & $10.71 \pm 0.006$ & $ 1.05 \pm 0.01 $ & $2.12 \pm 0.01$ & $1.07 \pm 0.007$ \\[.1ex]
\hline
\end{tabular}
\label{unsaturated}
\end{table*}

The magnitudes were calculated in the
VEGA photometric system 
using the zero points given in the WFPC2 handbook \citep{2004WFPC2}.
 Table~\ref{unsaturated} shows that the
stellar brightness differs between two epochs by $\approx$0.2\,mag.  The
second-epoch observations suggest that the source was brighter than  during the
first epoch.
The table also provides the stellar colour indices for comparison to those for the outer arc derived below. 

\subsection{Photometry on outer dust arc}
We measured the dust arc surface brightness at three different aperture locations 
in F606W-filter wide-field images from both epochs. The comparison is shown in Table~\ref{dust_color}.

\begin{table}[ht]
\caption{Magnitude per square arcsecond values for three aperture positions in the outer dust arc from the two epochs  of F606W images 
(see Fig. 7 as reference for the aperture locations).}.
\centering
\begin{tabular}{c|c c}
\hline \hline
& \multicolumn{2}{c}{Surface brightness [mag]} \\[0.5ex]
\cline{2-3}
F606W apertures & first epoch & second epoch \\[0.5ex]
\hline
1 & $21.77 \pm 0.20$ & $21.90 \pm 0.22$ \\
2 & $21.55 \pm 0.18$ & $21.62 \pm 0.19$ \\
3 & $21.41 \pm 0.15$ & $21.46 \pm 0.17$ \\[.1ex]
\hline
\end{tabular}
\label{dust_color}
\end{table}

The consistent brightness differences between the two epochs of 0.5$\dots$0.13\,mag for the three filters suggest that the outer arc became fainter. Although the photometric errors preclude a firm statement on variability, it seems certain that the arc did not brighten, unlike the star. This contradiction can be resolved by taking the light travel time from the star to the dust arc into account, which amounts to $\approx$45 days for the mean radial  distance of $\approx$8100\,au derived from the projected angular separation. It is even longer
if the arc is in the foreground. 
Keeping in mind the erratic short-term photometric variability of V1331 Cyg \citep{1983SvAL....9..289K,2012A&A...548A..79A}, which possibly results from accretion fluctuations, the difference in star and arc brightness trends is acceptable.

\subsection{Outer arc reddening analysis}
\begin{figure}[h]
\resizebox{\hsize}{!}{
\includegraphics[trim={1.7cm 3cm 1.3cm 4cm},clip]{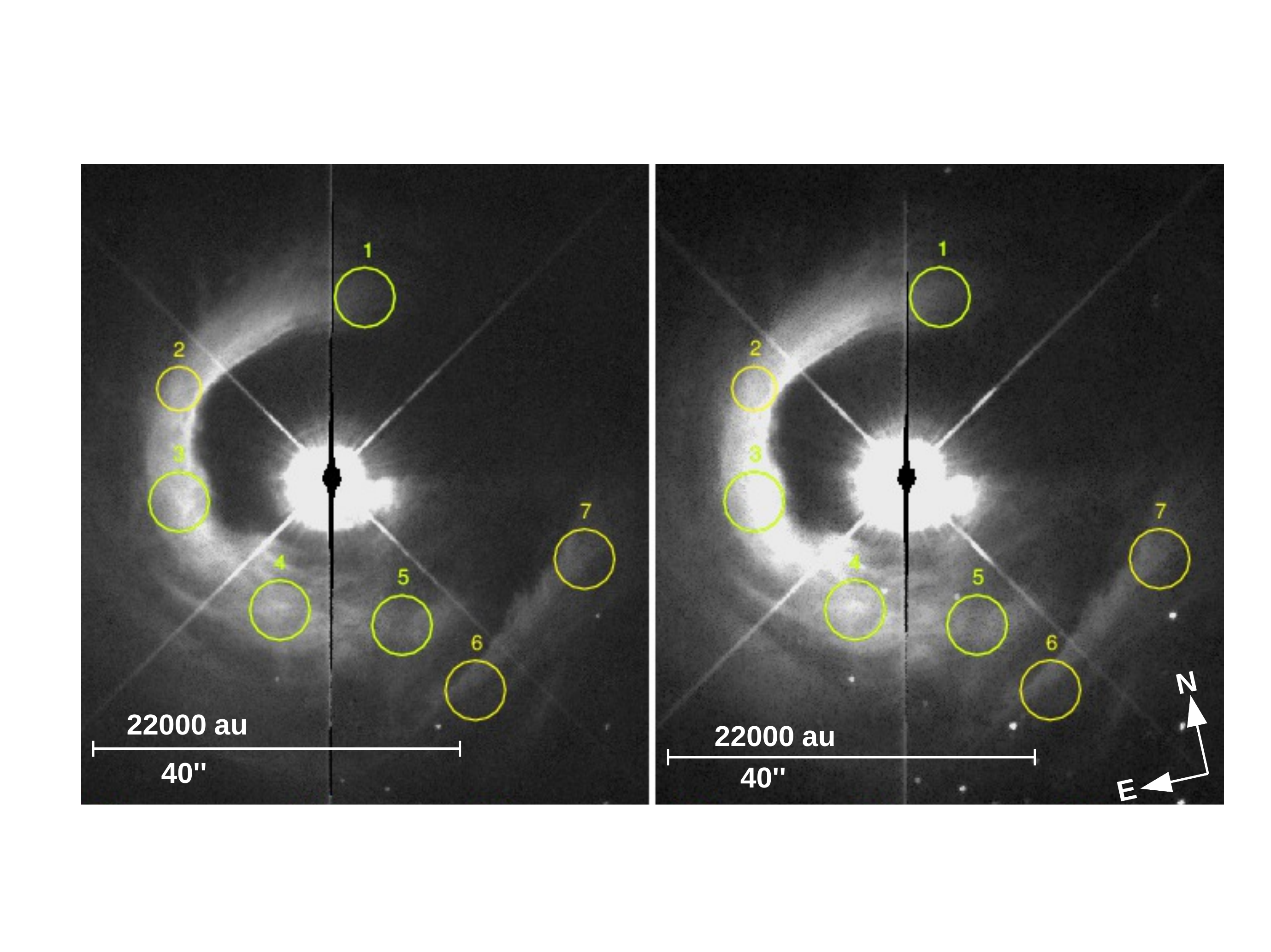}
}
\caption{Second-epoch F606W image (left) and F814W image (right) showing the apertures used to calculate the reddening. The same apertures were used to calculate the F450W colour.}
\label{red_Fig}
\end{figure}

The second-epoch data were also used to check for any signature of reddening in the northern part of the outer dust arc as compared to the rest that would be caused by the filament of dark cloud LDN 981. We calculated [F450W-F814W] and [F606W-F814W] values at seven different aperture positions located on the outer dust arc and the south-west wing. Table~\ref{red_table} shows details of  these calculations. Figure~\ref{red_Fig} shows the positioning of apertures as given in the table.

\begin{table*}[!ht]
\caption{Brightness (mag\,arcsec$^{-2}$) measured in all three filters in second epoch and the colour index calculated from it.}
\centering
\begin{tabular}{c|c c c|c c c}
\hline \hline
& \multicolumn{3}{c}{Surface brightness [mag]} & \multicolumn{3}{c}{Colour index [mag]} \\[0.5ex]
\cline{2-7}
Aperture & F450W & F606W & F814W & F450W-F606W & F450W-F814W & F606W-F814W \\
\hline
1 & $29.21 \pm 0.55$ & $27.68 \pm 0.18$ & $26.68 \pm 0.20$ & $1.53 \pm 0.57$ & $2.53 \pm 0.58$ & $1.00 \pm 0.26$ \\
2 & $27.91 \pm 0.45$ & $26.60 \pm 0.16$ & $25.67 \pm 0.18$ & $1.31 \pm 0.47$ & $2.24 \pm 0.48$ & $0.93 \pm 0.24$ \\
3 & $27.66 \pm 0.29$ & $26.38 \pm 0.11$ & $25.44 \pm 0.13$ & $1.28 \pm 0.31$ & $2.22 \pm 0.32$ & $0.94 \pm 0.17$ \\
4 & $28.09 \pm 0.32$ & $26.70 \pm 0.12$ & $25.76 \pm 0.14$ & $1.39 \pm 0.34$ & $2.33 \pm 0.35$ & $0.94 \pm 0.18$ \\
5 & $28.92 \pm 0.48$ & $27.58 \pm 0.18$ & $26.67 \pm 0.20$ & $1.34 \pm 0.51$ & $2.25 \pm 0.52$ & $0.91 \pm 0.26$ \\
6 & $29.75 \pm 0.75$ & $28.31 \pm 0.29$ & $27.36 \pm 0.32$ & $1.44 \pm 0.80$ & $2.39 \pm 0.81$ & $0.95 \pm 0.43$ \\
7 & $29.36 \pm 0.70$ & $28.03 \pm 0.26$ & $27.17 \pm 0.28$ & $1.33 \pm 0.74$ & $2.19 \pm 0.75$ & $0.86 \pm 0.38$ \\[.1ex]
\hline
\end{tabular}
\label{red_table}
\end{table*}

The observed  [F450W-F814W] colour index value of 2.53\,mag for aperture 1 clearly indicates some sort of reddening in this part of the arc.

\begin{figure}[h]
\includegraphics[width=9.1cm]{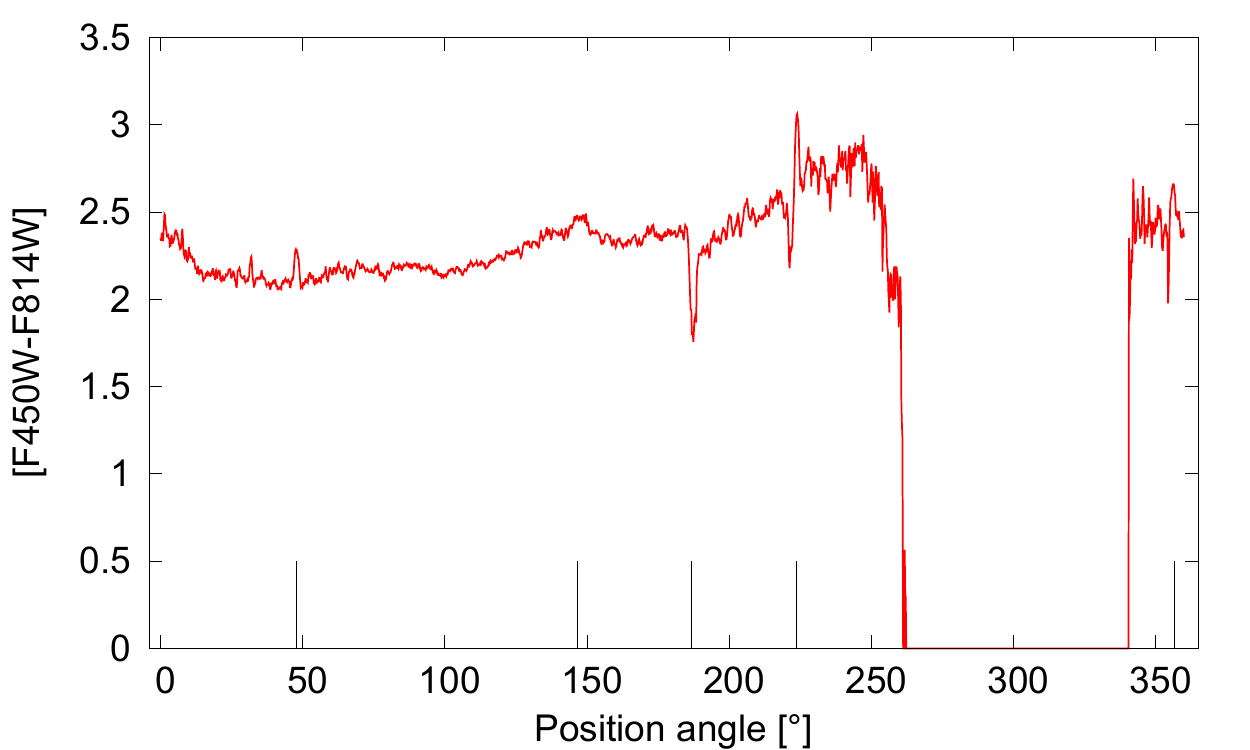}
\caption{Second-epoch outer dust arc azimuthal median profile for [F450W-F814W] colour. 
Vertical bars mark artefacts.
}
\label{median_x}
\end{figure}

To 
investigate whether the missing ring section might be (at least partially) due to extinction from the dark cloud, we calculated the azimuthal median profile based on the [F450W-F814W] colour image (see Fig.~\ref{f4_8}).
The method remained the same as detailed in Sect.~\ref{radial_analysis}, except that this time the azimuthal profile was obtained as the
{\em \textup{radial}} median 
for pixels within the annulus marked by the blue circles
over all position angles (PAs). The 
resulting plot is shown in Fig.~\ref{median_x}. The black vertical lines denote peaks caused by diffraction spikes 
and stellar bleeding on the CCD.

\subsection{Outer arc colour-brightness relation}

\begin{figure}[h]
\includegraphics[width=9cm]{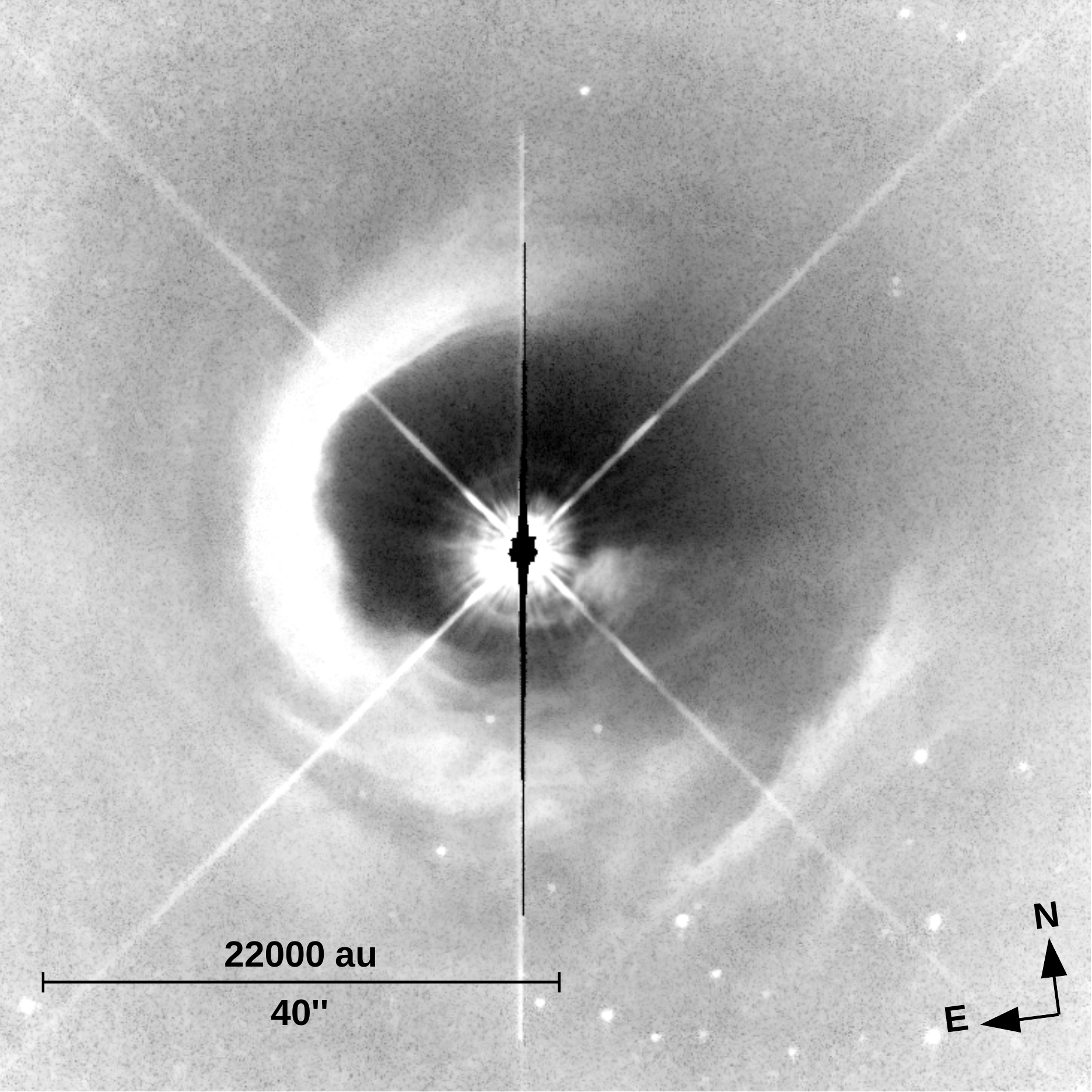}
\caption{
F814W image with radial flux weighting.
}
\label{re-scaled}
\end{figure}

\begin{figure}[h]
\includegraphics[trim={2cm .9cm 1.5cm 1.5cm},clip,width=9cm]{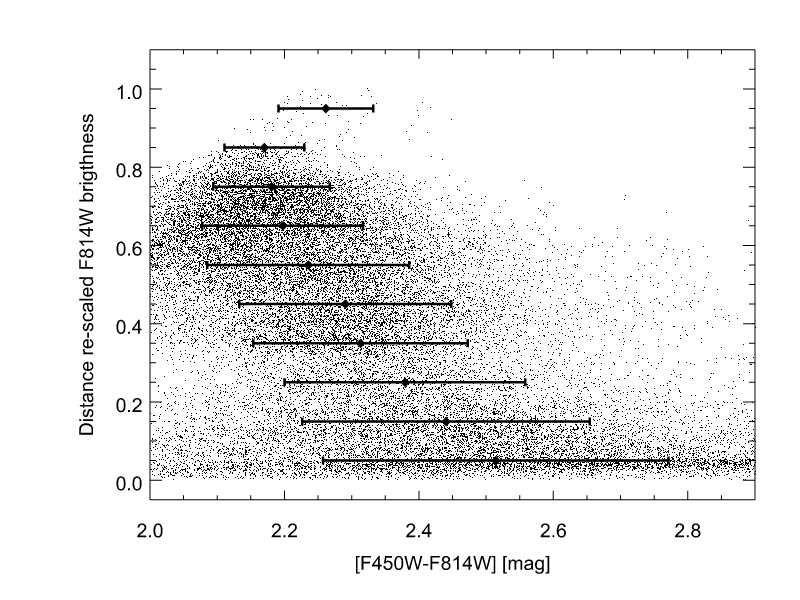}
\caption{Re-scaled brightness vs. colour plot for pixels of the outer arc confined by the annulus  shown in Fig.\,\ref{f4_8}.}
\label{pixels}
\end{figure}

Figures \ref{f4_8} and \ref{median_x} both indicate quite a range of the [F450W-F814W] colour and an increase over a PA range from about 30\degr$\dots$250\degr. While the colour variation might in principle result from spatial grain size variations, we consider this to be unlikely for the investigated region.  Under the assumption of a unitary size distribution, the azimuthal colour variation can be interpreted to be due to scattering. Since the scattering efficiency of dust grains depends on both wavelength and scattering angle, see \cite{2003ApJ...598.1017D},
 the scattered light will be bluer and brighter for small scattering angles (forward scattering). 
It has to be taken into account, however, that the light source V1331 Cyg is slightly displaced from the centre of the arc. This
means that the amount of incident radiation along the arc
depends on the position angle. The implied distance variation can be accounted for by assuming that the decrease of surface brightness with distance follows that of other reflection nebulae. \cite{1983AJ.....88..418C} found that this is approximately proportional to the inverse distance. Multiplying the brightness value of all pixel values by their respective distance (measured in the sky plane) from V1331 Cyg therefore virtually shifts them to ``unit'' distance from the star. Remarkably, the re-scaled F814W image shown in Fig.\,\ref{re-scaled} reveals the inner arcs identified by \cite{2007ApJ...656..287Q} even without PSF subtraction. The plot of the re-scaled brightness vs. colour index (Fig.\,\ref{pixels}) for each pixel in the annular region shown in Fig.\,\ref{f4_8} is displayed in Fig.\,\ref{pixels} along with the mean and standard deviation of the colour for brightness bins of 0.1 arbitrary units. Despite a certain fraction of outliers, bluer pixels are brighter as a general trend. We take this as evidence that dust grain scattering governs the colour and brightness distribution of the outer arc. 

We note that the brightness correction based on the projected distance is the conservative case. The areas of strong forward scattering (blue) have to be in the foreground, implying that their true distance from the star is larger. Consequently, the proper distance correction would make them even brighter as they appear in Fig.\,\ref{re-scaled} and enhance the trend seen in Fig.\,\ref{pixels}.

\subsection{Gap extinction analysis using UKIDSS images}
\begin{figure*}[htbp]
\centering
\includegraphics[trim={5cm 3.5cm 4cm 2cm},clip,width=8.7cm]{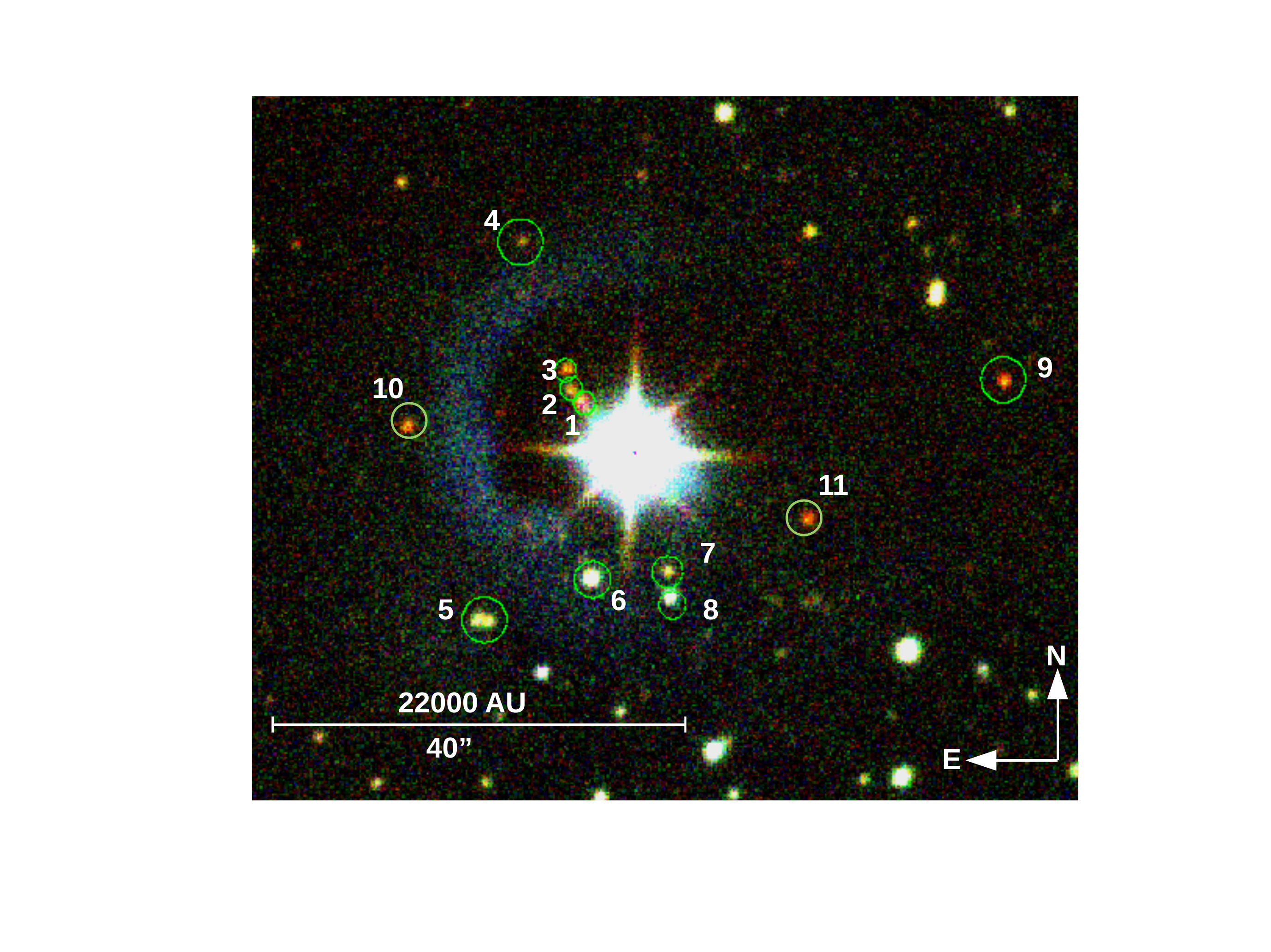}
\includegraphics[width=9.6cm]{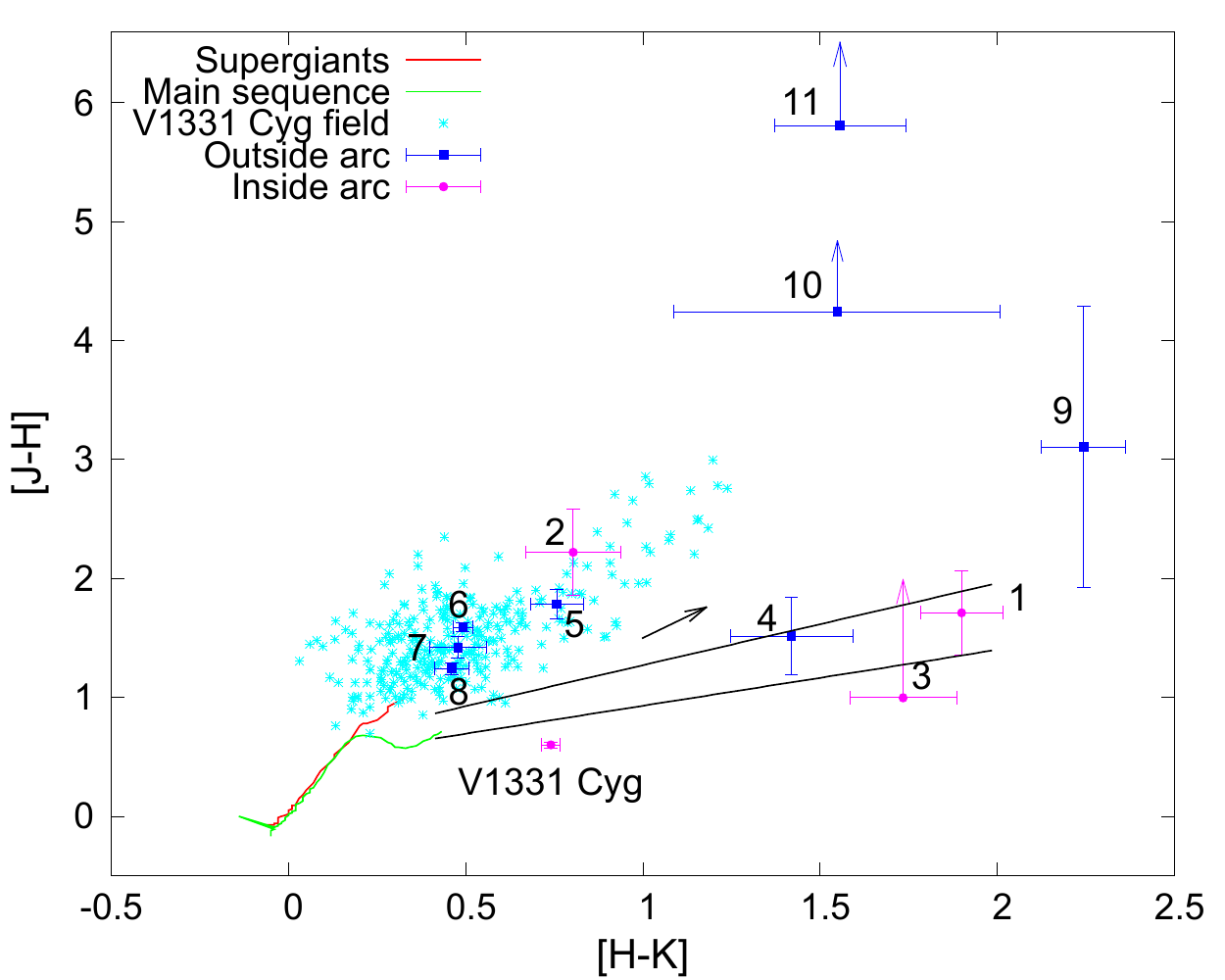}
\caption{Left: UKIDSS JHK-band colour image showing the stars marked for the reddening analysis. Right: The colour-colour plot for UKIDSS stars surrounding V1331 Cyg along with the reddening vector 
for $A_V$=2.4\,mag.
The two inclined lines mark the borders of dereddened classical T Tauri stars \citep{1997AJ....114..288M}.
The supergiants and main-sequence star data are adapted from \citet{1983A&A...128...84K}.}
\label{ukidss}
\end{figure*}

Figure~\ref{color} shows a gap between the outer and inner ring 
that is devoid of stars. The question therefore arises whether this space is truly empty, or if has such a high column density that it precludes the optical view on background stars.
The near-infrared UKIDSS images, on the other hand, show some faint stars in the gap between the outer ring and the star (see  Fig.~\ref{ukidss}). We compared the reddening of these stars to those in the environment to probe for dense matter that might
be associated with the protostellar environment and/or the molecular cloud.

We performed photometry on the separate near-infrared (NIR) images using SExtractor \citep{1996A&AS..117..393B}. The stars labelled 1, 2, and 3 in Fig.~\ref{ukidss} are the stars inside the gap, and stars  4 to 11 lie just outside the outer dust arc. 
The  proximity of star 1 to V1331 Cyg means that its photometry will be affected by the bright star. To avoid this, another star, similar in brightness to V1331 Cyg, was selected from the NIR images and was subtracted from the YSO  to cancel its scattering halo.
The  magnitudes for star 1 were calculated afterwards. Since SExtractor was unable to detect stars 3, 10, and 11 in J band, we used the limiting J-band magnitude to determine a lower bound on their [J-H] colour. The detection limit in J was derived from the magnitude statistics for completeness exceeding 50\%. The final colour-colour plot is shown in Fig.~\ref{ukidss}. Additional data for supergiant branch and main-sequence stars was used as a reference to overall photometry of V1331 Cyg field. The reddening vector corresponding to $A_V$=2.4\,mag was calculated using methods detailed in \citet{2005ApJ...619..931I}.


Of the stars within the gap, star 1 is possibly a candidate T Tauri star. Star 2 follows the reddening path for main-sequence stars. It seems to be extincted by $A_V$ of 7$\dots$10\,mag, but is still in the range of the adjacent field stars. Since there is only a lower limit on [J-H] for star 3, it could be a more heavily reddened star or a young stellar object showing infrared excess. The latter is indicated for the outer star 4 and possibly for star 9 as well. Stars 5 to 8 are mildly reddened objects, while stars 10 and 11 are strongly reddened. From this analysis we conclude that there  is no evidence for the stars in the gap to be more extincted than the surrounding stars. This suggests that there is no matter from the protostellar environment between the outer arc and the star, meaning that the gap is real.

\begin{figure}[h]
\includegraphics[trim={0.25cm 0.05cm 0.25cm 0.25cm},clip,width=9cm]{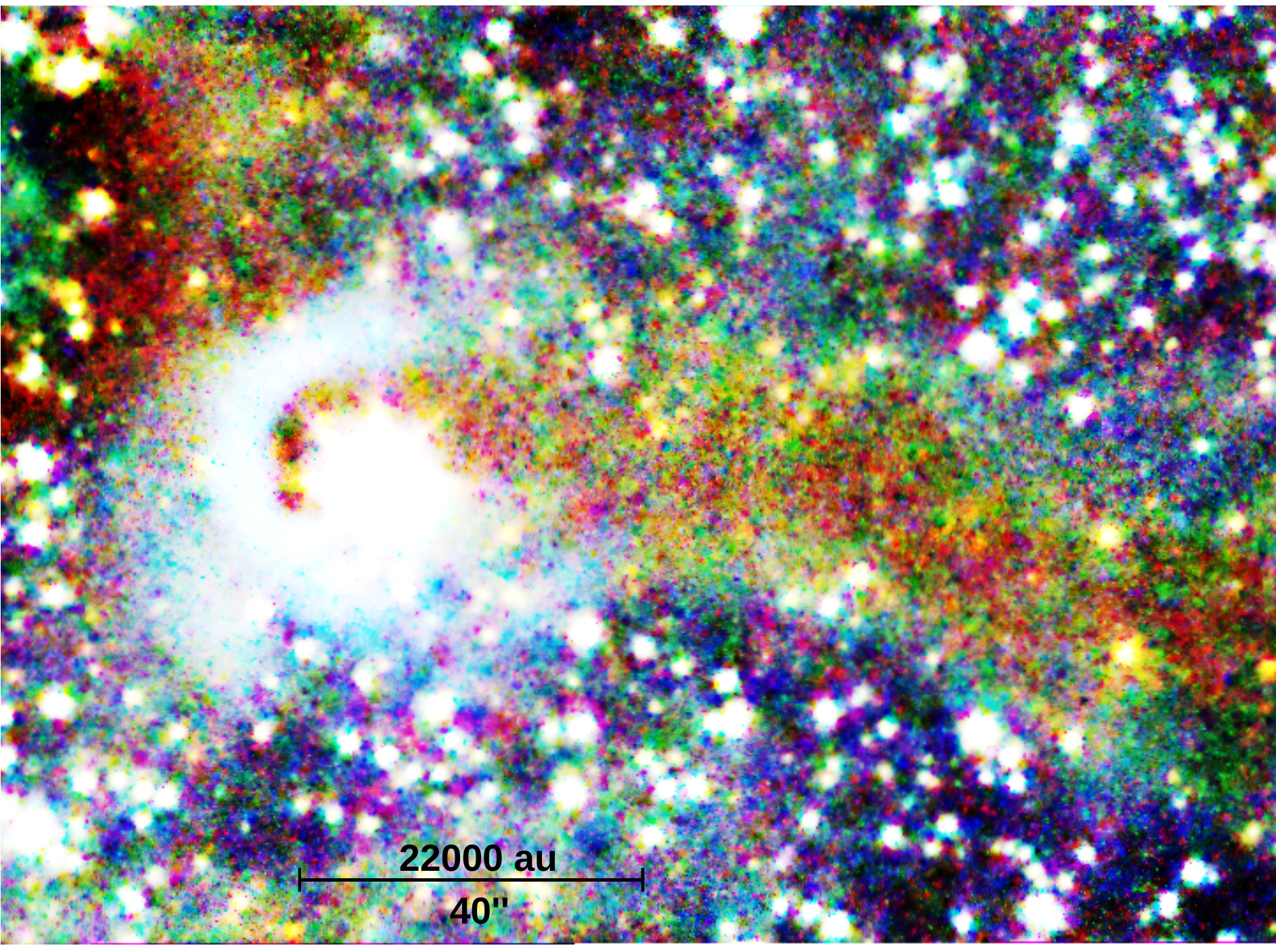}
\caption{
Histogram-equalised colour image based on adaptively smoothed JHK UKIDSS frames. V1331 Cyg  is in front of the bending filamentary dark cloud. The dark area at the north-east border is an image artefact.
}
\label{smooth_Fig}
\end{figure}

In addition to this particular 
analysis, the UKIDSS images also allow us to investigate
to which extent the dark cloud influences the appearance of the scattering nebula. Even with
the fidelity of HST images,
the strong scattering at optical wavelengths complicates discerning the relation between  LDN\,981 and the YSO. Since the smaller scattering cross-section of the dust grains at near-infrared wavelengths leads to a lower surface brightness, the  dark cloud morphology can be studied closer to the young star. To this aim, the images were adaptively smoothed to enhance low-level surface brightness using the algorithm of \citet{1996A&AS..118..575P}. The result shown in Fig.~\ref{smooth_Fig} reveals 
light scattered off the surface of the filamentary dark cloud, the so-called cloudshine 
(cf. \citealt{2006ApJ...636L.105F}). LDN\,981 stretches from west to east \textup{\textup{{\em \textup{behind}} }}V1331 Cyg, and turns north slightly east of the star. The dark cloud morphology corresponds to what has been seen in emission at coarse resolution with 
SCUBA \citep{2001ApJS..134..115S}
and {\sl HERSCHEL}/SPIRE \citep{2013ApJ...772..117G}.

This conclusion is supported by the reddening of stars next to V1331 Cyg (see Fig.~\ref{ukidss}). Stars 4 and 9 to 11 are much redder than stars 5 to 8 south of the YSO.

\subsection{HERSCHEL/PACS results}\label{Subsec:HERSCHEL-res}
Our attempt to detect extended thermal dust emission from V1331~Cyg is based on PSF subtraction. From the variety of target images and PSFs established with different data-processing techniques, we produced PSF-subtracted frames and searched for the one with the lowest residuals in the least-squares sense for each band. The {\sc JScanam} images with a pixel fraction of 0.1 applied in the dFrizzling algorithm in combination with non-drizzled PSFs based on re-centred Vesta observations performed on operational day 345 yielded the lowest chi-square values. That these PSFs are best suited for the drizzled target images may be explained by a contribution from warm dust of the inner rings, which slightly widens the PSF of V1331~Cyg. The {\sc PACS} PSFs were scaled to match the target peak flux value. 
An oversampling factor of three and  fractional pixel shifts were employed in the least-squares minimisation.
The two 160\,$\mu$m images obtained in parallel with the 70 and 100\,$\mu$m observations were registered separately with respect to the PSF before averaging them and performing the subtraction.

\begin{figure}[h]
\centering
\includegraphics[clip,width=9cm]{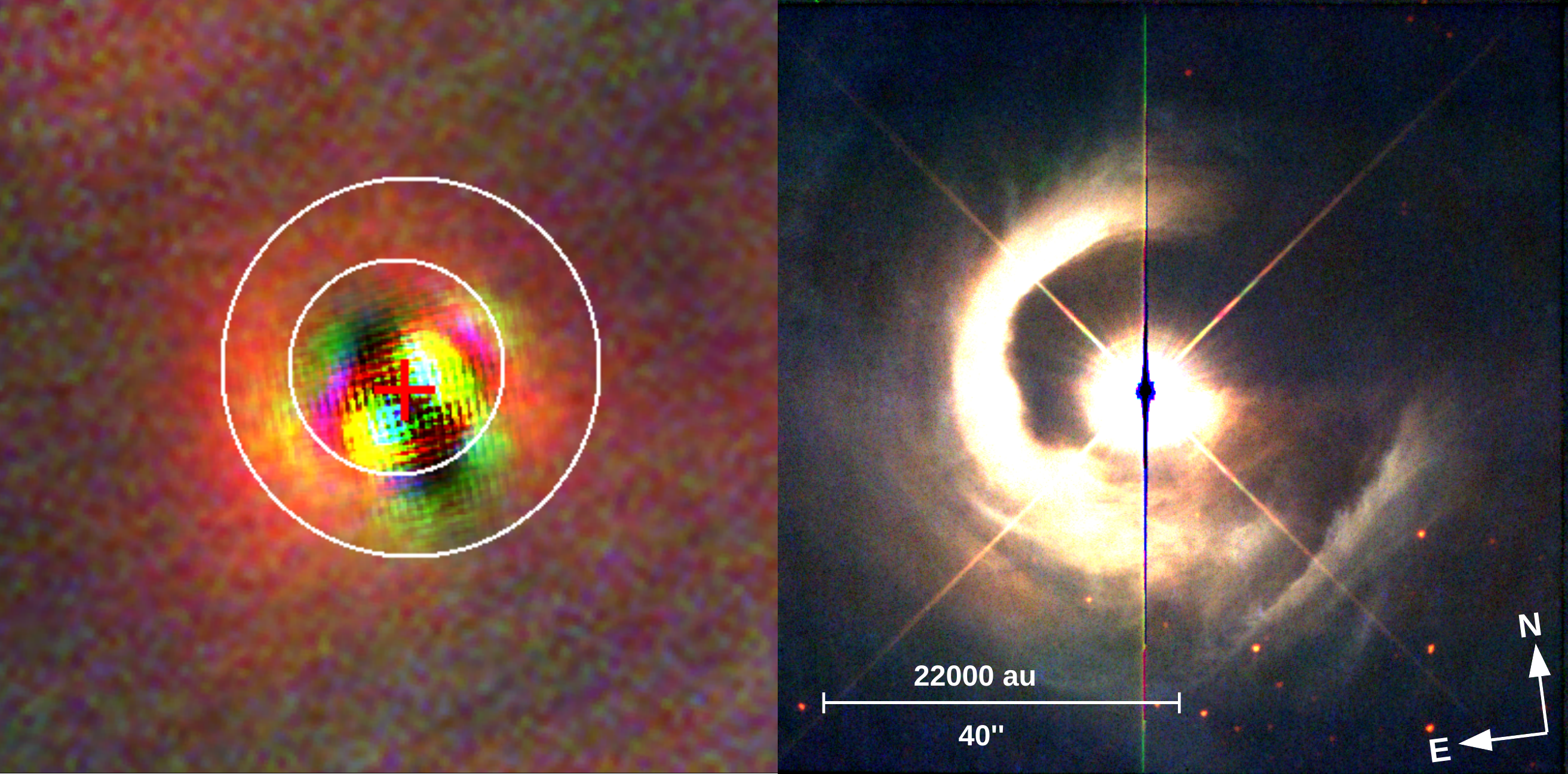}
\caption{{\sl HERSCHEL} colour-composite image 
of the residuals after PSF subtraction
(left) with the annular region showing emission from the outer arc, along with a HST second-epoch colour image (right) for comparison.}
\label{her_Fig}
\end{figure}

A colour-composite image based on the three PSF-subtracted frames is shown in Fig.~\ref{her_Fig}, where the white circles mark the boundaries of the outer annular scattering nebula. Since the residuals of the 70\,$\mu$m image are almost fully confined by the inner circle, we conclude that these are entirely due to PSF mismatch. For the 100\,$\mu$m channel, however, some residual emission is present within the confining circles, predominantly to the SW. Thus we tentatively conclude that it originates from the dust of the reflection nebula. This is supported by the clear detection of residual emission in this area at 160~$\mu$m, which even stretches to larger radii. Whether this extent is primarily due to the wavelength-dependent width of the PSF or a radial temperature gradient cannot be distinguished with the present data.

To estimate the 160\,$\mu$m flux, the residual emission was integrated within aperture radii of 11\arcsec{} and 35\arcsec{} centred on the star, taking the local background into account. This yielded a value of 1.06$\pm$0.01\,Jy. The formal error from the image noise is certainly only a lower bound. That the reflection nebula diminishes in the NIR (cf. Fig.~\ref{ukidss} left) indicates that it is certainly optically thin at the wavelengths observed with PACS. With this proposition and assumptions on the dust temperature, opacity and distance, a mass estimate can be derived. To this aim, we followed the approach by \citet{1994ApJ...433..157H} and used the same dust parameters. Since the tentative detection at 100\,$\mu$m cannot be used to infer a temperature, we adopted a value of 15\,K caused by the heating of V1331 Cyg and the interstellar radiation field. This yields a dust mass of $4.5\,M_{Earth}$. Of course, the mass estimate is very sensitive to temperature and drops to $1.0\,M_{Earth}$ for 20\,K.

\subsection{HHO spectroscopy}
\begin{figure}[h]
\centering
\includegraphics[trim={1.5cm 0.5cm 0.5cm 1cm},clip,width=9cm,height=6cm]{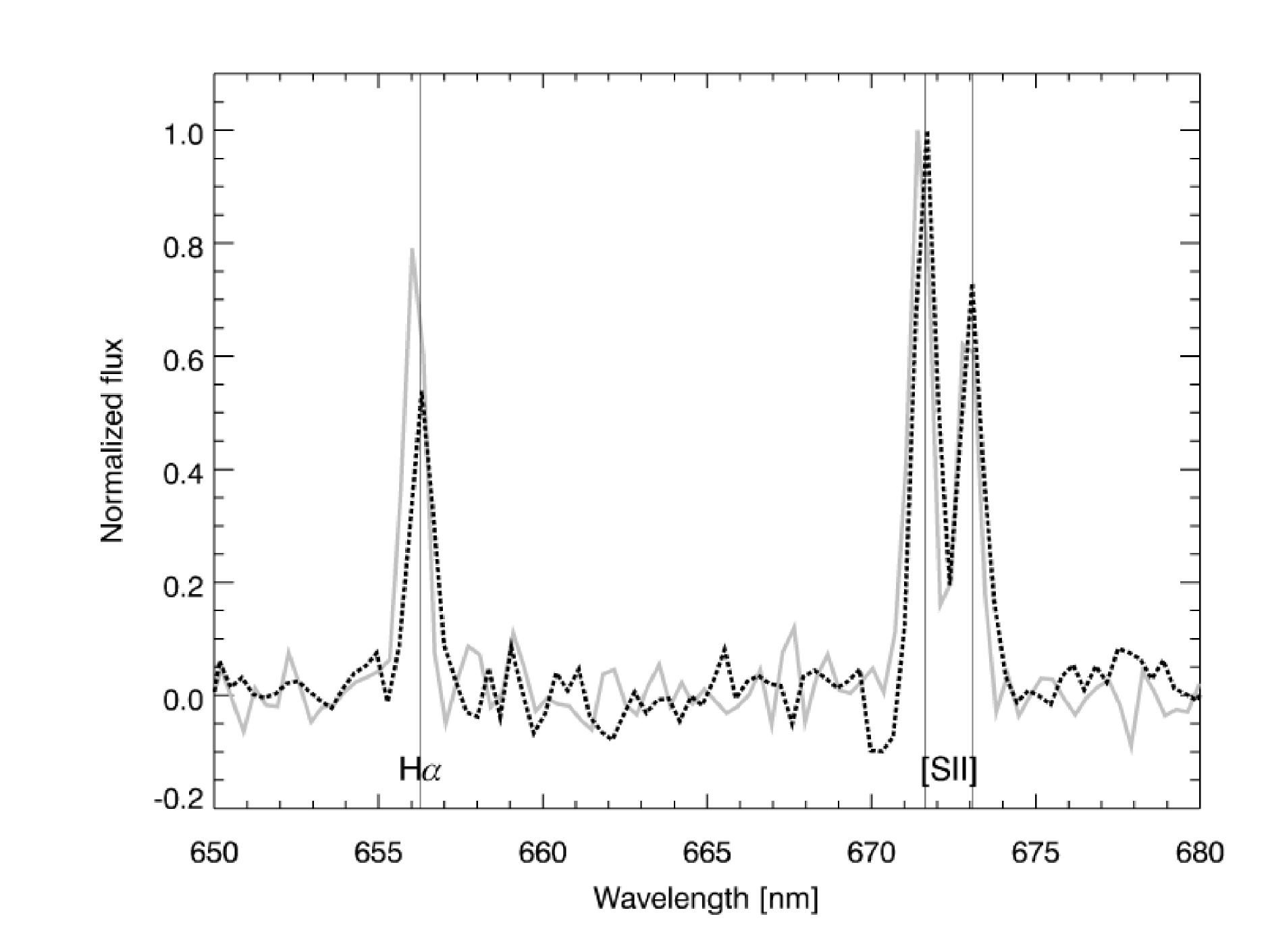}
\caption{Northern blueshifted (grey) and southern redshifted (black) HHO spectra obtained using the 2m Tautenburg telescope. The vertical grey lines denote the rest wavelengths of  H$\alpha$ and [SII] lines.}
\label{hhspec}
\end{figure}

The HHO spectra are shown in Fig.~\ref{hhspec}. Lines of H$\alpha$ (656.28 nm) and [SII] (671.7 nm and 673 nm) were detected for both northern and southern HH objects. The northern HHO is found to be blueshifted, while the southern HHO is redshifted. The peak signal-to-noise ratio for the southern HHO is 24 and about the same for the northern\ HHO. The radial velocities are 
$+$28\,km\,s$^{-1}$ for the southern HHO and $-$50\,km\,s$^{-1}$
for the northern HHO. The formal error of the velocity is about 10\,km\,s$^{-1}$. 
No other lines except those reported here were detected above the 3$\sigma$ level in the spectral range from 620\,nm to 745\,nm. 
The sense of the HH flow corresponds to that of the CO outflow claimed by \citet{1993AJ....106.2477M}. The northern HHO shows [OI] and [CaII] (marginal), which is not seen in the spectrum of the southern HHO. This is clear evidence for different excitation conditions. 


\section{Discussion}
\subsection{Dust arcs: Possible origin scenario}

Since we traced a slight expansion in the inner and outer dust arcs, there has to be a mechanism driving it.
It is reasonable to assume that V1331 Cyg originated from a dark cloud. Thus, the protostellar wind might have shaped the dust arcs we see today.
The star is seen almost pole-on, and the 
remains
of a blueshifted outflow cavity projected onto the sphere makes this a tempting explanation. 
However, several aspects argue against this.
Neither the decreasing outflow strength nor the widening of the opening angle over time can hardly produce the observed sharp structural features. Moreover, the blown-away dust will reside at a substantial distance from the star, implying a narrow range of the scattering angle, which cannot account for the azimuthal colour variation.

Another possibility is a past eruption as indicated by the results of \citet{1993AJ....106.2477M}.
The arcs would then result from the material blown out by an outburst, and thus are moving away from the star. \citet{1993AJ....106.2477M} modelled the dense CO gas  as an inclined torus around V1331 Cyg that expands with a overall velocity of 22$\pm$4\,km\,s$^{-1}$, which implies an implantation time of $\approx$4000$\pm 730$ years at the current expansion rate. The outer dust arc has an expansion velocity of $\approx$14.8$ \pm$ 3.6\,km\,s$^{-1}$. When we trace this velocity back for the mean radial distance of 8100\,au, it dates back to $\approx$2600$\pm 630$ years. With the error margin, this result seems to be consistent with the estimate made by \citet{1993AJ....106.2477M}. 

Similarly, tracing the inner arc expansion velocity back for the average radial distance of 3300\,au, the expansion would have started $\approx$10000$\pm$5000 years ago. 
At first glance, this result appears contradictory since the outer arc seemed to have overtaken the inner arcs during the expansion. However, this might result from  assuming a linear expansion velocity, which may not be valid if the velocity speeds up since the density decreases outward, for instance.  Last but not least, we note that relating the tangential velocity derived from the dust arc expansion to the overall velocity of the CO ring would yield the torus inclination. However, since the velocities were derived for different distances from the star, we refrain from doing so.


\subsection{Outer arc 
brightness and
colour variation: Possible reasons}

The outer dust arc shows both brightness and colour variations when considered over the azimuthal range (see Sect.~\ref{variable}). If the arc 
were a full annulus inclined with respect to the plane of the sky,
then the scattered light as seen in the dust arc would have had varying brightness distribution
depending upon the inclination angle. At inclination $\theta \geq 30^\circ $ , the 
facing side of the arc will brighten notably because forward scattering is more enhanced than backward scattering and vice versa.

The azimuthal colour and brightness trends  indicate systematic changes of scattering angle if a single grain-size distribution persists throughout the arc.  Indeed, our analysis of the re-scaled arc F814W brightness and the [F540W-F814W] colour supports this view. 
With this in mind, a more valid approximation for the outer arc structure could instead be a 3D spiral shape. Its NE part is
most likely farther away 
in the foreground, as indicated by the blue colour of the arc, resulting from a larger scattering angle. It winds towards south over east, and down closer to the star, thus getting slightly redder because the scattering angle decreases.
The SW part of the outer arc does not have well-defined inner boundary (see Fig.~\ref{color}). This could be the part of spiral turning inwards, being a possible remnant of past inflow,
and causing the colour variation. 
Interestingly, at the top of this region (west of the star), inner disk features were detected \citep{2009AIPC.1158..135K,2013prpl.conf2B063C}, which could be related to the accretion funnel. 
Similar large-scale inflowing spiral structures were recently detected with ALMA by \citet{2014ApJ...789L...4T}. 

Complementary high-resolution imaging polarimetry will help to disentangle the 3D structure of the environment of V1331 Cyg by using the dependence between polarisation degree and scattering angle.

\subsection{SW 
ridge
and elliptical arc}

The SW ridge, 
which is located
at $\approx$13000\,au 
projected
distance from the star, does not seem to be connected to the
outer dust arc. 
The [F450W-F814W] colour index image (see Fig.~\ref{f4_8}) shows that it is quite blue at a uniform level. This suggests that the SW ridge is in the foreground and unrelated to the expanding outer arc. From the difference image (see Fig.~\ref{diff_im}) we found evidence for motion towards V1331 Cyg.

\begin{figure}
\resizebox{\hsize}{!}{
\includegraphics[trim={1.3cm 2.3cm 1.3cm 2.3cm},clip]{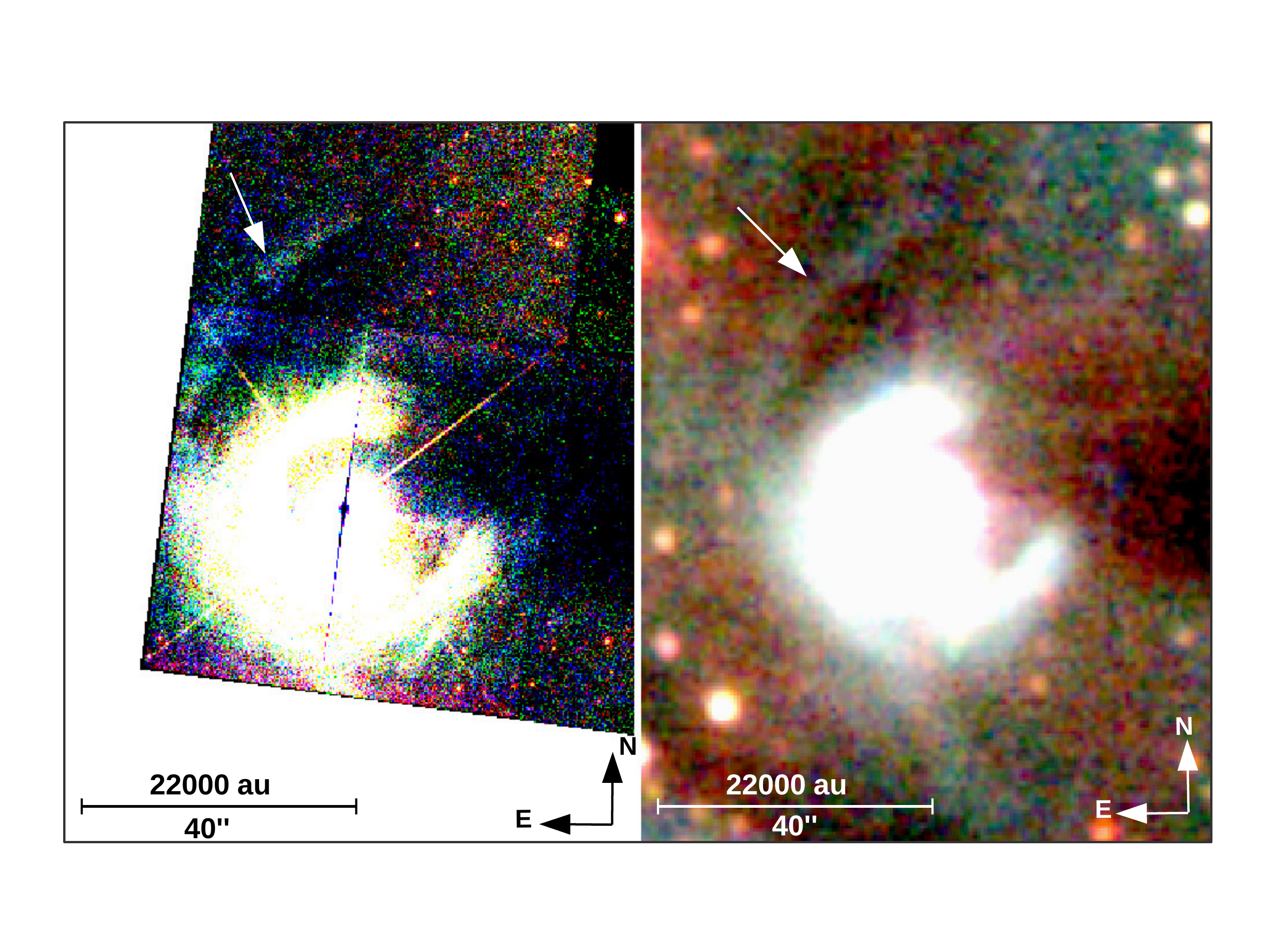}
}
\caption{HST second-epoch colour image (left) compared with a TLS image (right). The newly found arc-like feature is marked in both images.}
\label{SW1}
\end{figure}

In the HST second-epoch (Fig.~\ref{SW1}) image, we observe another faint arc-like feature from NE to north, lying farther outwards than the outer dust arc. This feature, unlike the outer dust arc, is more elliptical in shape. When seen up close in the Digital Sky survey (DSS) and the TLS image, this feature appears to be linked with the SW ridge in some way.  However, this cannot be justified, simply because there is also a gap between the newly found arc and the SW ridge, and the connecting turn is also very sharp in angle.


\subsection{What causes the missing arc section?} \label{ldn}

\begin{figure*}[t]
\centering
\includegraphics[width=15cm]{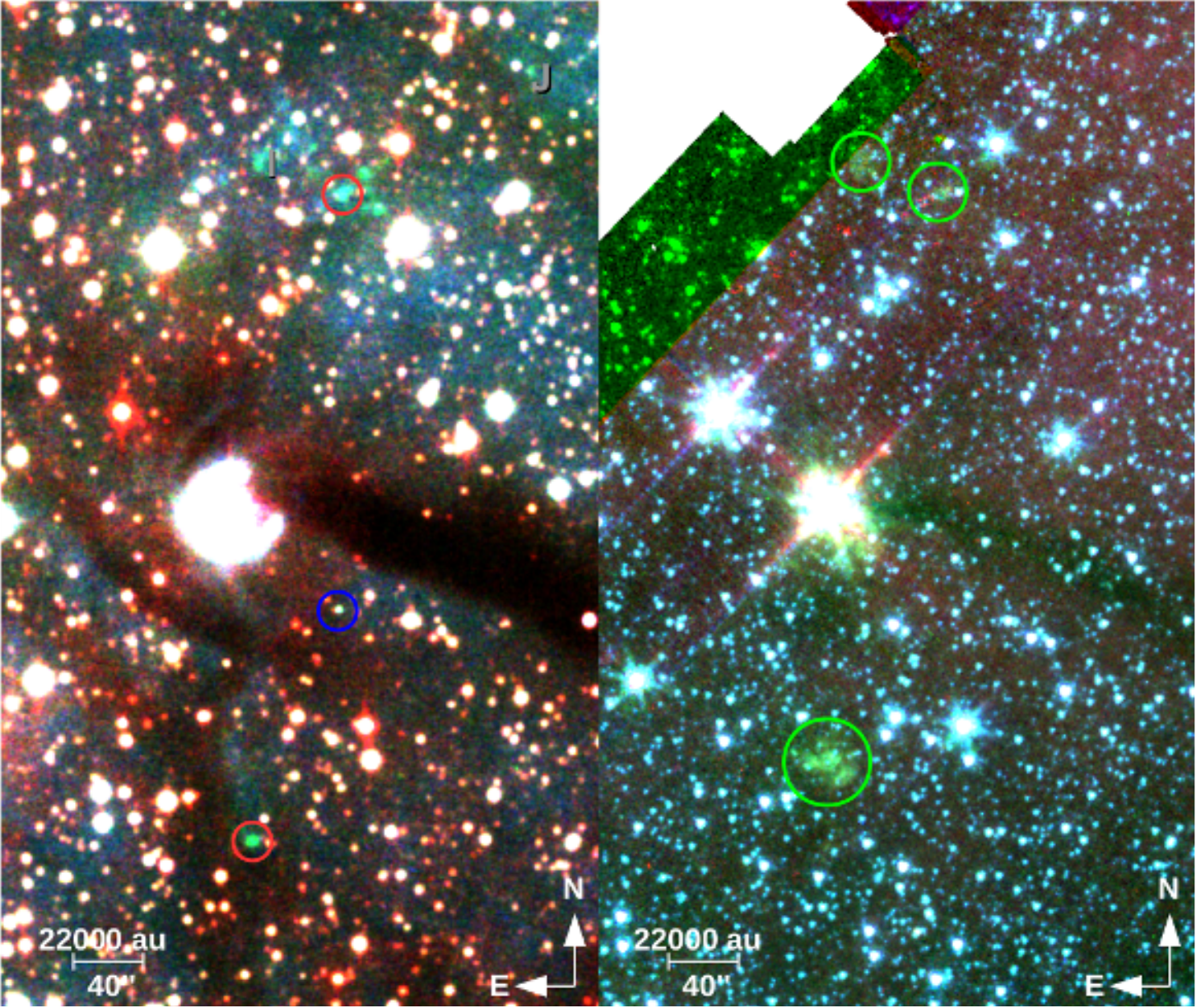}
\caption{TLS image (left) and SPITZER-IRAC (right) showing the bipolar outflow from V1331 Cyg. The optical RGB image (left) is composed of I, H$\alpha$, and [SII] frames. The newly found HHOs I and J to the north are labelled according to the nomenclature of \citet{1998AJ....116..860M}. The two brightest HHOs  for which spectra were obtained are encircled in red. The blue circle marks an emission-line star without notable infrared excess. The 4.5\,$\mu$m emission from shocked H$_2$ is marked by green circles in the IRAC image (right).
}
\label{hhflow}
\end{figure*}

With the proposition that the outer arc is a full annulus, the question arises as to the cause of the missing section. It is tempting to assume that the dark cloud blocks the light. Since there is a gradient of the column density within the dark cloud towards its border,  the width of the obscured section will depend on wavelength simply because the dark cloud appears to look
slimmer
in the red than in the blue.  This would cause a reddening at the edge zones of the missing section. This has been found to
be the case with the photometry of region 1 (see Fig.~\ref{red_Fig}) and in the azimuthal profile (see Fig.~\ref{median_x}) at PAs of $\approx$250\degr and $\approx$340\degr). On the other hand, the extinction analysis of the UKIDSS images showed that the cloud is located behind the star. These two findings can only be reconciled if the dark cloud is in the foreground of V1331 Cyg in the west and runs behind the YSO towards the east. But to obscure the stellar environment, it then has to come close to the star. This implies, however, that the dark cloud would also scatter stellar light and appear in reflection. The deep TLS image (see Fig.~\ref{hhflow}) shows some scattering in the I band that might originate from the stellar illumination of the dark cloud. Its weak surface brightness indicates, however, that the LDN\,981 cannot be very close to the star.

Another viable explanation is shadowing. Since the mutual location of the circumstellar disk of V1331 Cyg and the scattering grains is unknown, it cannot be ruled out that the dust of the outer arc resides in an elliptical torus that is offset and oblique with respect to the circumstellar disk plane. If this is the
case, the disk could cast a shadow onto the torus that appears as the missing section. The azimuthal variation of the shadow size follows the same principle as for the dark cloud consideration. This configuration is supported by the model for the CO ring of \cite{1993AJ....106.2477M}, which is inclined by 30\degr{} to the plane of the sky. The ring that confines the outer arc is blueshifted to the east and redshifted to the west of V1331Cyg. If the dust is moving in the same fashion as the CO, it will be in the foreground of the star in the east while in the background to the west. The implied wide range of the scattering angle might explain the overall brightness and colour variations. However, some regions remain problematic, such as the arc section to the north, where the colour decreases very strongly (see Figs.\,\ref{f4_8}
and \ref{median_x}).

A similar case, albeit on a smaller spatial scale, might be HD\,135344B, for which shadowing of the outer disk by a misaligned inner dust belt has recently been claimed  \citep{2016arXiv...1603...00481}.

Shadowing might also occur as a result of height variations in the circumstellar disk.
For YSOs seen close to edge-on, this causes photometric changes known as UX Ori variability, see \citet{1994AJ....108.1906H}. The first-epoch PC frames indicate a compact feature very close to the star \citep{2013prpl.conf2B063C} at the correct position angle that is connected to a helical arm in the outer disk
found by \cite{2009AIPC.1158..135K}. It might be a scale-height enhancement in the outer circumstellar disk that stretches over some azimuthal range. However, its orbital timescale is by far
too long to observe changes in PA of the shadow over the epoch difference of our HST observations.
Moreover, shadowing will also lead to a lower dust temperature and in turn to reduced thermal dust emission, which is along the lines of the PACS observations. Given the observational evidence, we consider shadowing as a likely explanation for the missing arc section. 




\subsection{Jet, Herbig-Haro objects, and bipolar outflow}

V1331 Cyg is associated with a bipolar outflow, as shown in Fig.~\ref{hhflow}. With a viewing angle of between $30\degr\dots40$\degr, the bipolar outflow was estimated previously to have an extent of $\approx$1\,pc. Now, with the recent evidence of viewing the star almost pole-on 
\citep{2011ApJ...738..112D,2014MNRAS.442.3643P}, the bipolar flow length
could be of 10\,pc or even more. Since the small inclination implies a large uncertainty of the outflow length, we estimate the kinematic age of the flow. Moreover, the relatively low radial velocity for HHOs that are observed almost head-on might indicate their deceleration along the path.

The flow appears to be slightly bent instead of being straight. 
This deflection might be due to the interaction with surrounding material that diverts the path of the outflow. Another possible explanation would be precession induced by the binarity of the driving source. The faint spiral-like emission feature stretching to the south from V1331 Cyg might indicate binarity if it were a wiggling jet, as has been claimed by \citet{1998AJ....116..860M}. However, the deep TLS image shows that the wiggling structure instead represents the photo-ionised skin of a filamentary molecular cloud. It seems possible that the redshifted flow interacts with this cloud, giving rise to the bright southern HHO. The obvious difference in the radial velocities of the HHOs is not extraordinary. Asymmetric jets have been observed frequently (for example DG Tau). 

 \citet{1998AJ....116..860M} considered the possibility that the HHOs to the north-west (D, E, and F in their nomenclature) might be part of another flow since the HHOs depart from a straight flow geometry. We found a nearby flow from an embedded YSO on SPITZER-IRAC images that  does not cross the flow of V1331 Cyg,
however. The detection of an even more western HHO (J) confirms that all the patchy northern HHOs belong to the blueshifted flow.

%

The speckled 
morphology of the shock-excited
emission in optical lines and in molecular hydrogen does not resemble bow shocks at all. 
According to the models of  \citet{2010A&A...513A...5G}, 
it suggests a near to head-on view onto the outflow, 
in agreement with existing evidence.

\section{Conclusions}
The main conclusions of our study are as follows.
The proper motion analysis of HST/WFPC2 imaging for V1331 Cyg with an epoch difference of almost ten years yielded an average expansion velocity of $\approx$14.8$\pm$3.6\,km\,s$^{-1}$ for the outer arc and between 1.0\,km\,s$^{-1}$ and 3.0\,km\,s$^{-1}$ for the inner arc. 
The velocity for the outer arc results in trace-back times that
are consistent with the CO torus  implantation time. This supports the view that V1331 Cyg has undergone an outburst a few thousand years ago, and strengthens evidence that the YSO is the only verified {\it \textup{post}}-FUOR known to date.

The analysis of the overall brightness and colour variation of the outer arc indicates that they result from azimuthal variation of the scattering angle. Possible dust configurations could be a tilted misaligned torus that is confined by the CO ring of \cite{1993AJ....106.2477M} or a spiral structure that could be an accretion flow relic that tightens and winds down to the outer circumstellar disk. The option of the origin of dust arcs from the bipolar flow was considered but dismissed for various reasons, owing to the large span of the flow among others.

Evidence was found that the missing arc section is likely caused by shadowing from the circumstellar disk and is not due to extinction from LDN\,981. 


While we were unable to confirm the particle separation hypothesis based on the colour analysis of the outer dust arc, evidence for a larger portion of smaller grains at distances exceeding 15000\,au was found.

By viewing the star almost pole-on, the length of HH flow might be 10 parsec or even more. 

\begin{acknowledgements}
A.Ch. acknowledges valuable input and partly supervision from Karl Stapelfeldt (NASA/GSCF) and the feedback from colleagues at the TLS, received in preparation and during her PhD defense.
Helpful discussions with Jochen Eislöffel are acknowledged. The HST general observer program-11976 is based in part on observations made with the NASA/ESA Hubble Space Telescope, and obtained from the Hubble Legacy Archive, which is a collaboration between the Space Telescope Science Institute (STScI/NASA), the Space Telescope European Coordinating Facility (ST-ECF/ESA) and the Canadian Astronomy Data Centre (CADC/NRC/CSA).
This research has made use of the NASA/ IPAC Infrared Science Archive, which is operated by the Jet Propulsion Laboratory, California Institute of Technology, under contract with the National Aeronautics and Space Administration.
This research has made use of the SIMBAD database, operated at CDS, Strasbourg, France.
This research has made use of the VizieR catalogue access tool, CDS, Strasbourg, France. The original description of the VizieR service was published in 
\citet{2000A&AS..143...23O}.
\end{acknowledgements}

\end{document}